\documentclass[%
  aip,jcp,
 amsmath,amssymb,
 reprint,%
]{revtex4-1}

\usepackage{graphicx}
\usepackage{epstopdf}
\usepackage{dcolumn}
\usepackage{bm}
\usepackage{verbatim}

\usepackage[utf8]{inputenc}
\usepackage[T1]{fontenc}
\usepackage{mathptmx}

\begin{document}


\title[Concerted Hydration Free Energy]{Alchemical Transformations for Concerted Hydration Free Energy Estimation with Explicit Solvation}

\author{Sheenam Khuttan$^\ast$}
\altaffiliation{Ph.D. Program in Biochemistry, The Graduate Center of the City University of New York, New York, NY}
\author{Solmaz Azimi$^\ast$}
\altaffiliation{Ph.D. Program in Biochemistry, The Graduate Center of the City University of New York, New York, NY}
\thanks{These two authors contributed equally to the work}
\author{Joe Z. Wu}
\altaffiliation{Ph.D. Program in Chemistry, The Graduate Center of the City University of New York, New York, NY}
\author{Emilio Gallicchio}
\altaffiliation{Ph.D. Program in Biochemistry, The Graduate Center of the City University of New York, New York, NY}
\altaffiliation{Ph.D. Program in Chemistry, The Graduate Center of the City University of New York, New York, NY}
\email[Corresponding author: ]{egallicchio@brooklyn.cuny.edu}
\affiliation{Department of Chemistry, Brooklyn College of the City University of New York, New York, NY. \\ $^\ast$These authors contributed equally to this work.}

\date{\today}

\begin{abstract}

  We present a family of alchemical perturbation potentials that enable the calculation of hydration free energies of small to medium-sized molecules in a concerted single alchemical coupling step instead of the commonly used sequence of two distinct coupling steps for Lennard-Jones and electrostatic interactions. The perturbation potentials are based on the softplus function of the solute-solvent interaction energy designed to focus sampling near entropic bottlenecks along the alchemical pathway. We present a general framework to optimize the parameters of alchemical perturbation potentials of this kind. The optimization procedure is based on the $\lambda$-function formalism and the maximum-likelihood parameter estimation procedure we developed earlier to avoid the occurrence of multi-modal distributions of the coupling energy along the alchemical path. A novel soft-core function applied to the overall solute-solvent interaction energy rather than  individual interatomic pair potentials critical for this result is also presented. Because it does not require modifications of core force and energy routines, the soft-core formulation can be easily deployed in molecular dynamics simulation codes.  We illustrate the method by applying it to the estimation of the hydration free energy in water droplets of compounds of varying size and complexity. In each case, we show that convergence of the hydration free energy is achieved rapidly. This work paves the way for the ongoing development of more streamlined algorithms to estimate free energies of molecular binding with explicit solvation. 

\end{abstract}

\maketitle

\section{\label{sec:intro}Introduction}

The hydration free energy of a compound is defined as the reversible work for transferring one molecule of the compound from the gas phase to the water phase.\cite{Ben-Naim:74} The hydration free energy is an important characteristic of a substance. For instance, it is one of the determinant factors of the water solubility of drug formulations\cite{bergstrom2018computational} and of the binding affinity of an inhibitor towards a protein receptor.\cite{Gilson:Given:Bush:McCammon:97} Hydration free energies are most commonly derived from Henry's constant measurements.\cite{mobley2014freesolv}

Free energies of hydration can also be estimated computationally.\cite{Chipot:Pohorille:book:2007} Most commonly, this is accomplished by molecular simulations of alchemical transformations in which solute-solvent interactions are progressively turned on. The nature of the alchemical transformation is critical to obtaining reliable results. A single direct alchemical process in which the coupling between the solute and the solvent is increased in a simple linear fashion has been found to be problematic for anything other than the smallest solutes (monoatomic atoms and ions, water, methane, and similar).\cite{de2011free} One issue is  the singularity of the derivative of the alchemical potential with respect to the parameter $\lambda$ near the decoupled state.\cite{simonson1993free} Simple approaches of this kind can also be found to converge slowly due to bottlenecks along the alchemical path caused by poorly sampled conformational equilibria.\cite{mobley2012let}

These issues have been the subject of intense studies. This collective effort resulted in a set of recommended best practices for alchemical calculations commonly employed by the computational chemistry community and are nowadays implemented in popular molecular simulation software packages.\cite{Chipot:Pohorille:book:2007,klimovich2015guidelines,lee2020alchemical} For example, it is generally recommended to split the coupling of the solute and the solvent into two phases, the first in which the volume-exclusion core repulsion and dispersion interactions are turned on, followed by a second phase in which electrostatic interactions are turned on. useful for avoiding end-point singularities, especially when volume-exclusion core repulsion terms are introduced..\cite{Steinbrecher2007} In addition,  it is often necessary for large solutes to introduce one small group of atoms at a time or to resort to bond-growing processes in which solute atoms are pushed out from a central point rather than directly created in the solvent.\cite{pitera2002comparison}

While successful, these strategies add layers of complexity to hydration free energy calculations.\cite{lee2020improved} Splitting the coupling process into two calculation phases requires specifying two alchemical schedules that are often very different from each other. This practice also relies on simulating a nonphysical intermediate state consisting of the uncharged solute in water which could have very different conformational propensities than the actual solute. The unphysical uncharged solute could, for example, undergo hydrophobic collapse and revert slowly to the physical extended hydrated state. 

Software implementations of soft-core potential functions are cumbersome and difficult to maintain, and  they require the optimization of  specific parameters for each interaction potential type.\cite{lee2020improved} Conventional separation-shifted scaled soft-core pair potentials\cite{Steinbrecher2007}  also lead to non-linear and non-algebraic forms of the alchemical potential that require a cumbersome energy rescoring process for the estimation of free energy changes with thermodynamic reweighting.\cite{li2020repulsive,Shirts2008a}  

Thus, it would be beneficial to develop more streamlined alchemical protocols that (i) compute the hydration free energy in one continuous transformation process, (ii) do not require modifications of the standard form of pair potential functions,  and (iii) can be parameterized using a linear progression of the charging parameter $\lambda$, while (iv) preserving or improving the rate of equilibration and convergence of free energy estimates relative to mainstream protocols. Here we present a method with potentially has all of these characteristics while being applicable to a range of alchemical transformations, including those used in structure-based drug design\cite{cournia2020rigorous} that is our primary focus.\cite{bfzhang2016rnaseh,pal2019dopamine}

The approach presented here is based on the work we carried out recently to optimize binding free energy calculations in the context of implicit solvation.\cite{kilburg2018assessment,pal2019perturbation} This work yielded a family of optimized alchemical potential energy functions that modify all interactions in one concerted step and that do not require customized soft-core pair potentials. This effort also resulted in the development of a general framework, based on generalized non-Boltzmann sampling theory\cite{lu2014investigating} and maximum likelihood estimation,\cite{kilburg2018analytical} for the optimization of the parameters of the alchemical potentials to accelerate conformational mixing and convergence. 

In the present work, we demonstrate that the same approach applies to the estimation of the hydration free energies of small to medium-sized molecules using an explicit representation of the solvent. As in our previous work,\cite{pal2019perturbation} the approach is based on the observation that the slow convergence of free energies is due to rare conformational transitions caused by entropic bottlenecks akin to first-order phase transitions,\cite{kim2010generalized} and that these can be circumvented by an appropriate choice of the alchemical perturbation energy function.

This work is organized as follows. We first review the theoretical and computational protocol developed earlier. Then we apply it to a model system of hydration in which four diverse solutes of a wide range of sizes (ethanol, alanine dipeptide, 1-naphthol, and 3,4-diphenyltoluene) are transferred from the gas phase to a water droplet. We show that the optimized protocols yield results consistent with more conventional approaches when these can reach convergence. However, for larger solutes, only the optimized protocol can achieve rapid and reliable convergence. These promising outcomes pave the way for a novel generation of more efficient and more streamlined methodologies for alchemical transformations in condensed phases.  

\section{\label{sec:methods}Theory and Methods}

\subsection{Alchemical Transformations for the Estimation of Hydration Free Energies}

Alchemical transformations are based on an alchemical potential energy function that interpolates from the potential energy function $U_0$ of the starting state to that of the final state $U_1$ as the progress parameter $\lambda$ goes, conventionally, from $0$ to $1$. The most straightforward approach is a linear interpolating function of the form:
\begin{equation}
  U_{\lambda}(x)=U_{0}(x)+ \lambda [ U_1(x) - U_0(x) ] = U_0(x) + \lambda u(x)
  \label{eq:pert_pot_linear_simple}
\end{equation}
where $x$ represents the degrees of freedom of the system and we have introduced the perturbation function
\begin{equation}
u(x) =  U_{1}(x)-U_{0}(x) \label{eq:binding-energy}
\end{equation} 
which in the case of hydration corresponds to the solute-solvent interaction energy and, critical for the following developments, does not depend on $\lambda$. 

While straightforward, this simple alchemical potential energy function leads to instabilities and poor convergence, especially when volume-exclusion terms are introduced.\cite{lee2020improved} To address these issues while maintaining as much as possible the same simple framework, we consider the family of alchemical potential energy functions of the form:
\begin{equation}
U_{\lambda}(x)=U_{0}(x)+W_{\lambda}[u_{\rm sc}(x)]\label{eq:pert_pot}
\end{equation}
where $u_{\rm sc}(x)$, defined below, is a soft core-modified solute-solvent interaction energy, and $W_\lambda(u_{\rm sc})$ is an alchemical perturbation function defined such that $W_{0}(u_{\rm sc})=0$ and $W_{1}(u_{\rm sc})=u_{\rm sc}$ at $\lambda=0$ and $\lambda=1$, respectively.  We will consider in this work the linear function
\begin{equation}
W_{\lambda}(u_{\rm sc})=\lambda u_{\rm sc}
\label{eq:linear-function}
\end{equation}
and the generalized softplus function
\begin{equation}
  W_{\lambda}(u_{\rm sc})=\frac{\lambda_{2}-\lambda_{1}}{\alpha}\ln\left[1+e^{-\alpha(u_{\rm sc}-u_{0})}\right]+\lambda_{2}u_{\rm sc}+w_{0}
  \label{eq:ilog-function}
\end{equation}
where the parameters $\lambda_{2}$, $\lambda_{1}$, $\alpha$, $u_{0}$, and $w_{0}$ are functions of $\lambda$. The derivative of the softplus function is the generalized logistic function (Fermi's function):
\begin{equation}
  \frac{\partial W_{\lambda}(u_{\rm sc})}{\partial u_{\rm sc}}=\frac{\lambda_{2}-\lambda_{1}}{1+e^{-\alpha(u_{\rm sc}-u_{0})}}+\lambda_{1} \, .
  \label{eq:ilog-function-der}
\end{equation}
The softplus perturbation function's parameters are optimized using the procedure described in reference \onlinecite{pal2019perturbation} and briefly described later in this Section.

In this work, we employ the following definition of the soft-core solute-solvent interaction energy\cite{pal2019perturbation}
\begin{equation}
u_{{\rm sc}}(u)=\begin{cases}
u & u\le 0\\
u_{{\rm max}} f_\text{sc}(u/u_{{\rm max}}) & u>0
\end{cases}\label{eq:soft-core-general}
\end{equation}
where
\begin{equation}
f_\text{sc}(y) = \frac{z^{a}-1}{z^{a}+1} \label{eq:rat-sc} \, ,
\end{equation}
where $u_{{\rm max}} > 0$ is the maximum allowed value of the soft-core solute-solvent interaction energy,  $z=1+2 y/a + 2 (y/a)^2$ and $a$ is an adjustable dimensionless exponent (here $u_{\rm max} = 50$ kcal/mol and $a=1/16$). With these definitions, $u_{\rm sc}(u)$ is a $C(2)$-smooth one-to-one map from the original solute-solvent interaction energy to the soft-core interaction energy. For favorable solute-solvent interaction energies ($u<0$), the two are the same. However, for unfavorable solute-solvent interaction energies ($u>0$), such as when two atoms clash approaching each other, the soft-core interaction energy $u_{\rm sc}$ grows less rapidly than $u$, and it eventually reaches a maximum plateau, unlike the solute-solvent interaction energy that grows indefinitely.

We stress that in this soft-core methodology, there are no soft-core modifications of pair potentials. The soft-core function is applied to the overall solute-solvent interaction energy $u$ evaluated with the standard form of the Coulomb and Lennard-Jones interatomic potentials without soft-core modifications.

To obtain the hydration free energy, a set of samples of the soft-core solute-solvent interaction energies, $u_{\rm sc}(i)$, are collected during molecular dynamics simulations performed at a sequence of $\lambda$ values between $0$ and $1$. The free energy profile as a function of $\lambda$, $\Delta G(\lambda)$, is obtained by multi-state reweighting\cite{Shirts2008a} using the UWHAM method.\cite{Tan2012} The hydration free energy in the Ben-Naim standard state\cite{Gallicchio:Kubo:Levy:98} is by definition the value of free energy profile at $\lambda=1$, $\Delta G^{\ast}_h = \Delta G(1)$.

\subsection{Analytical Theory of Alchemical Transformations}

The hydration process is analyzed and optimized using the theory we recently developed for alchemical potential energy functions of the form of Eq.\ (\ref{eq:pert_pot}).\cite{kilburg2018analytical} Briefly, following the potential distribution theorem,\cite{PDTbook:2006} in the present case we consider  $p_0(u_{\rm sc})$, the probability density of the soft-core solute-solvent interaction energy $u_{\rm sc}$ in the ensemble in which solute where solute and solvent are uncoupled ($\lambda=0$). All other quantities of the alchemical transformation can be obtained from $p_0(u_{\rm sc})$.\cite{Gallicchio2010,Gallicchio2011adv} In particular, given $p_0(u_{\rm sc})$, the probability density for the binding energy $u_{\rm sc}$ for the state with perturbation potential $W_\lambda (u_{\rm sc})$ is 
\begin{equation}
p_{\lambda}(u_{\rm sc})=\frac{1}{K(\lambda)}p_{0}(u_{\rm sc}) \exp\left[-\beta W_\lambda (u_{\rm sc})\right]\label{eq:plambdau_1}
\end{equation}
where $\beta = 1/k_B T$,
\begin{equation}
K(\lambda)=\int_{-\infty}^{+\infty}p_{0}(u_{\rm sc}) \exp\left[-\beta W_\lambda (u_{\rm sc})\right] du_{\rm sc} =\langle \exp\left[-\beta W_\lambda (u_{\rm sc})\right] \rangle_{\lambda=0}\label{eq:plambdau_2}
\end{equation}
is the excess component of the equilibrium constant for binding and
\begin{equation}
\Delta G(\lambda) = - \frac{1}{\beta} \ln K(\lambda) \label{eq:gblambda}
\end{equation}
is the corresponding free energy profile. Note that for a linear perturbation potential, $W_\lambda(u_{\rm sc}) = \lambda u_{\rm sc}$, Eqs.~(\ref{eq:plambdau_2}) and (\ref{eq:gblambda}) state that the free energy profile is related to the double-sided Laplace transform of $p_\lambda(u_{\rm sc})$.

An analytical description of $p_0(u_{\rm sc})$, and thus of all the quantities derived from it, is available. Briefly (see reference \onlinecite{kilburg2018analytical} for the full derivation), the theory is based on the assumption that the statistics of, in this case, the solute-solvent interaction energy $u$ in the decoupled state is the convolution of two processes. One that described by the sum of many ``soft'' background solute-solvent interactions, and that follows central limit statistics and another process that follows max statistics that describes ``hard'' atomic collisions. 
The probability density $p_0(u)$ is further expressed as the superposition of probability densities of a small number of modes  
\begin{equation}
p_0(u) = \sum_i c_i p_{0i}(u) \label{eq:superp}
\end{equation}
where $c_i$ are adjustable statistical weights summing to $1$ and $p_{0i}(u)$ is the probability density corresponding to mode $i$ described analytically as (see reference \onlinecite{kilburg2018analytical} and appendix A of reference \onlinecite{pal2019perturbation} for the derivation):
\begin{eqnarray}
&&  p_{0i}(u) = p_{bi} g(u;\bar{u}_{bi},\sigma_{bi}) \nonumber \\
&& + (1-p_{bi}) \int_{0}^{+\infty}p_{WCA}(u';n_{li},\epsilon_{i},\tilde{u}_i)g(u-u';\bar{u}_{bi},\sigma_{bi})du' \label{eq:p0(u)conv2}
\end{eqnarray}
where $g(u;\bar{u},\sigma)$ is the normalized Gaussian density function of mean $\bar{u}$ and standard deviation $\sigma$ and
\begin{equation}
  p_{WCA}(u;n_l , \epsilon, \tilde{u}) = n_{l}\left[1-\frac{(1+x_{C})^{1/2}}{(1+x)^{1/2}}\right]^{n_{l}-1} 
  \frac{H(u)}{4\epsilon_{LJ}}\frac{(1+x_{C})^{1/2}}{x(1+x)^{3/2}}\, , \label{eq:pwcaf}
\end{equation}
where $x = \sqrt{u/\epsilon+\tilde{u}/\epsilon+1}$ and $x_C = \sqrt{\tilde{u}/\epsilon+1}$. The model for each mode $i$ depends on a number of adjustable parameters corresponding to the following physical quantities\cite{kilburg2018analytical}:
\begin{itemize}
\item $c_i$: statistical weight of mode $i$
\item $p_{bi}$: probability that no atomic clashes occur while in mode $i$
\item $\bar{u}_{bi}$: the average background interaction energy of mode $i$
\item $\sigma_{bi}$: the standard deviation of background interaction energy of mode $i$
\item $n_{li}$: the effective number of statistical uncorrelated atoms of the solute in  mode $i$
\item $\epsilon_{i}$: the effective $\epsilon$ parameter of an hypothetical Lennard-Jones interaction energy potential describing the solute-solvent interaction energy in mode $i$
\item $\tilde{u}_i$: the solute-solvent interaction energy value above which the collisional energy is not zero in mode $i$
\end{itemize}
The parameters above, together with the weights $c_i$, are varied to fit, using a maximum likelihood criterion, the distributions of the soft-core solute-solvent interaction energy obtained from numerical simulations.\cite{kilburg2018analytical} The distribution of the soft-core solute-solvent interaction energy  $p_0(u_{\rm sc})$ is obtained from $p_0(u)$ using the standard formula for the change of random variable.\cite{pal2019perturbation}   

In this work we use the analytical theory above and the results of trial alchemical simulations to optimize the parameters of the softplus alchemical perturbation potential in Eq.\ (\ref{eq:ilog-function}).\cite{pal2019perturbation} The procedure consists of running a trial alchemical calculations using the the linear alchemical potential $W_\lambda(u_{\rm sc}) = \lambda u_{\rm sc}$. The set of samples of the soft-core binding energies as a function of $\lambda$ obtained from the trial simulation are then used to derive optimized parameters for the analytical model for $p_0(u_{\rm sc})$ [Eqs.\ (\ref{eq:superp}) to (\ref{eq:pwcaf})] using a maximum likelihood approach.\cite{lee2012new} We developed an application based on the Tensorflow\cite{tensorflow2015-whitepaper} to conduct the maximum likelihood optimization ({\tt https://github.com/egallicc/femodel-tf-optimizer}).

The analytical form of $p_0(u_{\rm sc})$ so obtained is then analytically differentiated with respect to $u_{\rm sc}$ to obtain the so-called $\lambda$-function:\cite{pal2019perturbation}
\begin{equation}
\lambda_{0}(u_{\rm sc}) \equiv \frac{1}{\beta}\frac{\partial\ln p_{0}(u_{\rm sc})}{\partial u_{\rm sc}}\label{eq:lambda-function} \, .
\end{equation}
Minima and maxima of $p_\lambda(u_{\rm sc})$ occur when the $\lambda$-function and $\partial W_\lambda(u_{\rm sc})/\partial u_{\rm sc}$ intersect\cite{pal2019perturbation}
\begin{equation}
  \lambda_{0}(u_{\rm sc}) = \frac{\partial W_\lambda(u_{\rm sc})}{\partial u_{\rm sc}} \, ,
  \label{eq:lambda-function-intersection}
\end{equation}
Hence, the $\lambda$-function can be used as a guide to design alchemical potentials that avoid the occurrence of multi-modal distributions that are difficult to converge. The linear alchemical potential $W_\lambda(u_{\rm sc}) = \lambda u_{\rm sc} $ leads to $\partial W_\lambda(u_{\rm sc})/\partial u_{\rm sc} = \lambda = {\rm constant}$, which in many cases intersects the $\lambda$-function at multiple points. To avoid multi-modal distributions, here we use the softplus alchemical potential and we vary the parameters $\lambda_{2}$, $\lambda_{1}$, $\alpha$, $u_{0}$, and $w_{0}$ as a function of $\lambda$ such that the derivative of the softplus function in Eq.\ (\ref{eq:ilog-function-der}) intersects $\lambda_{0}(u_{\rm sc})$ at a single point at each $\lambda$ or, when this is not easily achievable, such that it at least intersects it at nearby points. As thoroughly discussed in reference \onlinecite{pal2019perturbation} this procedure removes or reduces the severity of entropic sampling bottlenecks during the alchemical coupling process and enhances conformational sampling efficiency and convergence of the binding free energy estimates.  

\subsection{Computational Details}

\begin{figure}
  \centering
  \includegraphics[scale = 0.35]{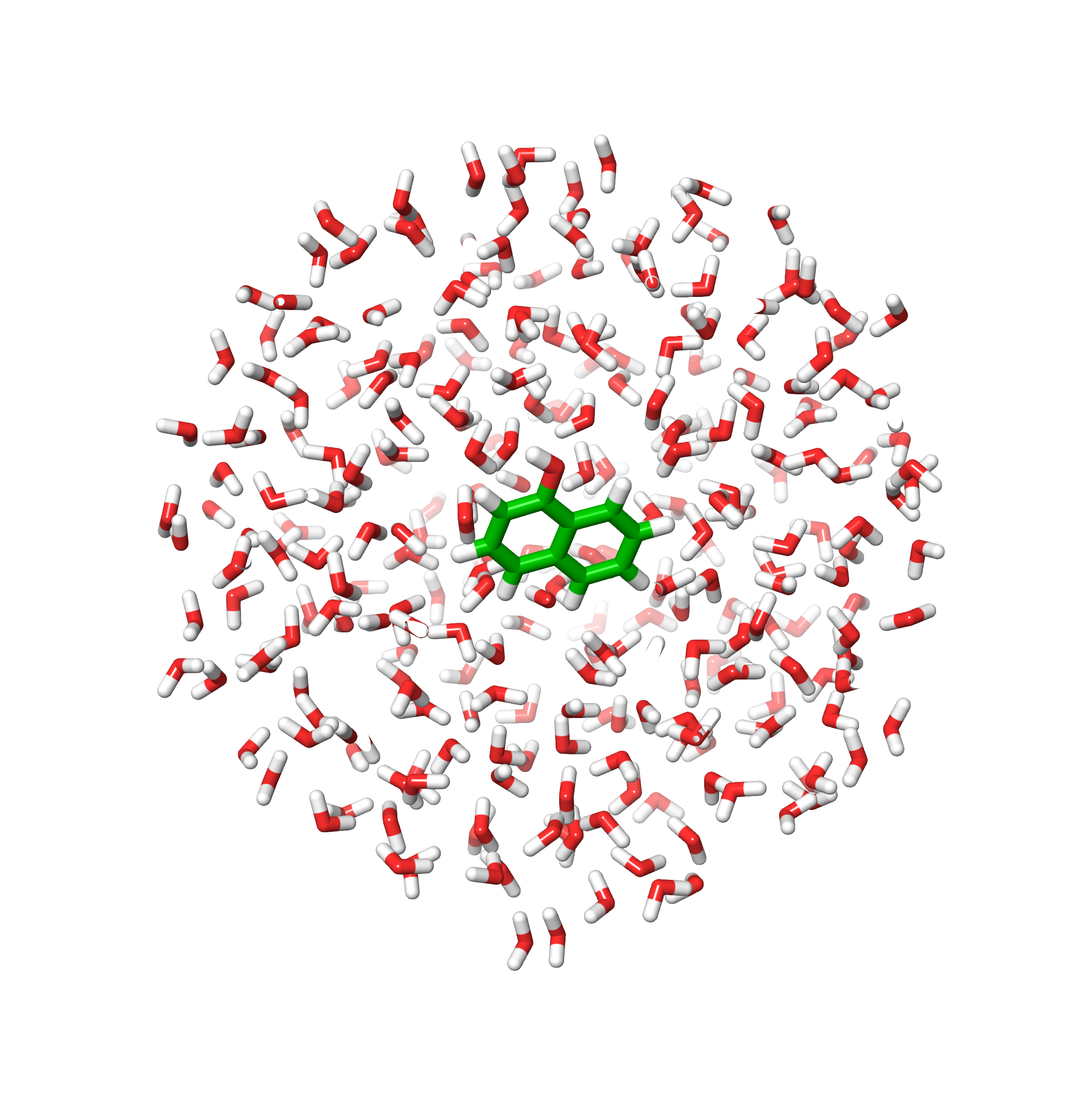}
  \caption{Illustration of 1-naphthol (green carbon atoms) inserted at the center of a droplet of water. Water molecules obscuring the solute are not shown for clarity.
    \label{fig:droplet}
  }
\end{figure}

In this work, we employ a hydration model in which solutes are transferred from vacuum to near the center of mass of a water droplet of about 27 \AA\ in diameter composed of 357 TIP3P water molecules (Figure \ref{fig:droplet}). The droplet is confined in a spherical region defined by a flat-bottom harmonic restraining potential centered at the origin and acting on each TIP3P water oxygen atom. The flat-bottom potential tolerance was set to 24 \AA\ resulting in a confinement region of approximately twice the droplet volume, and a force constant of 5 kcal/(mol \AA$^2$) beyond this tolerance. This simplified hydration model is selected here to illustrate and assess the proposed methodology using our existing SDM single-decoupling plugin in OpenMM,\cite{pal2019perturbation,eastman2017openmm} which was originally designed for binding free energy calculations with implicit solvation. As an added bonus, the droplet model adopted here avoids potential issues not relevant to this work concerning the correctness of implementations of alchemical  processes with periodic boundary conditions and long-range electrostatic.\cite{pan2017quantitative} 

All solutes were prepared with Maestro and GAFF/AM1-BCC force field parameters were assigned using the Antechamber program.\cite{wang2004development} Single Decoupling alchemical calculations were prepared using the SDM workflow ({\tt github.com/\-egallicc/openmm\_sdm\_worflow.git}) using 16 $\lambda$ steps. The MD calculations employed the OpenMM\cite{eastman2017openmm} MD engine the SDM integrator plugins ({\tt github.com/\-rajatkrpal/\-openmm\_sdm\_plugin.git}) using the OpenCL platform. The ASyncRE software,\cite{gallicchio2015asynchronous} customized for OpenMM and SDM ({\tt github.com/\-egallicc/\-async\_re-openmm.git}), was used for the Hamiltonian Replica Exchange in $\lambda$ space with an uniform $\lambda$ schedule between $0$ and $1$. The $\lambda$-dependent parameters used with the softplus potential are listed in Tables \ref{tab:ethanol-ilog-schedule} to \ref{tab:diphenyltoluene-ilog-schedule}. Molecular dynamics runs were conducted for a minimum of 5 ns per replica with a 1 fs time-step at $300$ K, exchanging $\lambda$ values approximately every 5 ps. A Langevin thermostat at $300$ K with a relaxation time constant of $20$ ps was used. Binding energy samples and trajectory frames were recorded every 5 ps. Calculations were performed on the XSEDE Comet GPU HPC cluster at the San Diego Supercomputing Center.

In this work, we employ the Hamiltonian Replica Exchange algorithm\cite{Sugita2000,Felts:Harano:Gallicchio:Levy:2004,Ravindranathan:Gallicchio:Levy:2006,Okumura2010} in alchemical space to accelerate conformational sampling.\cite{Woods2003,Rick2006,Gallicchio2010}
For the calculations reported here, we have employed the asynchronous implementation of Replica Exchange (ASyncRE)\cite{gallicchio2015asynchronous} with the Gibbs Independence Sampling algorithm for state reassignments.\cite{pal2019perturbation}

\begin{figure}
    \centering
    \includegraphics[scale = 0.35]{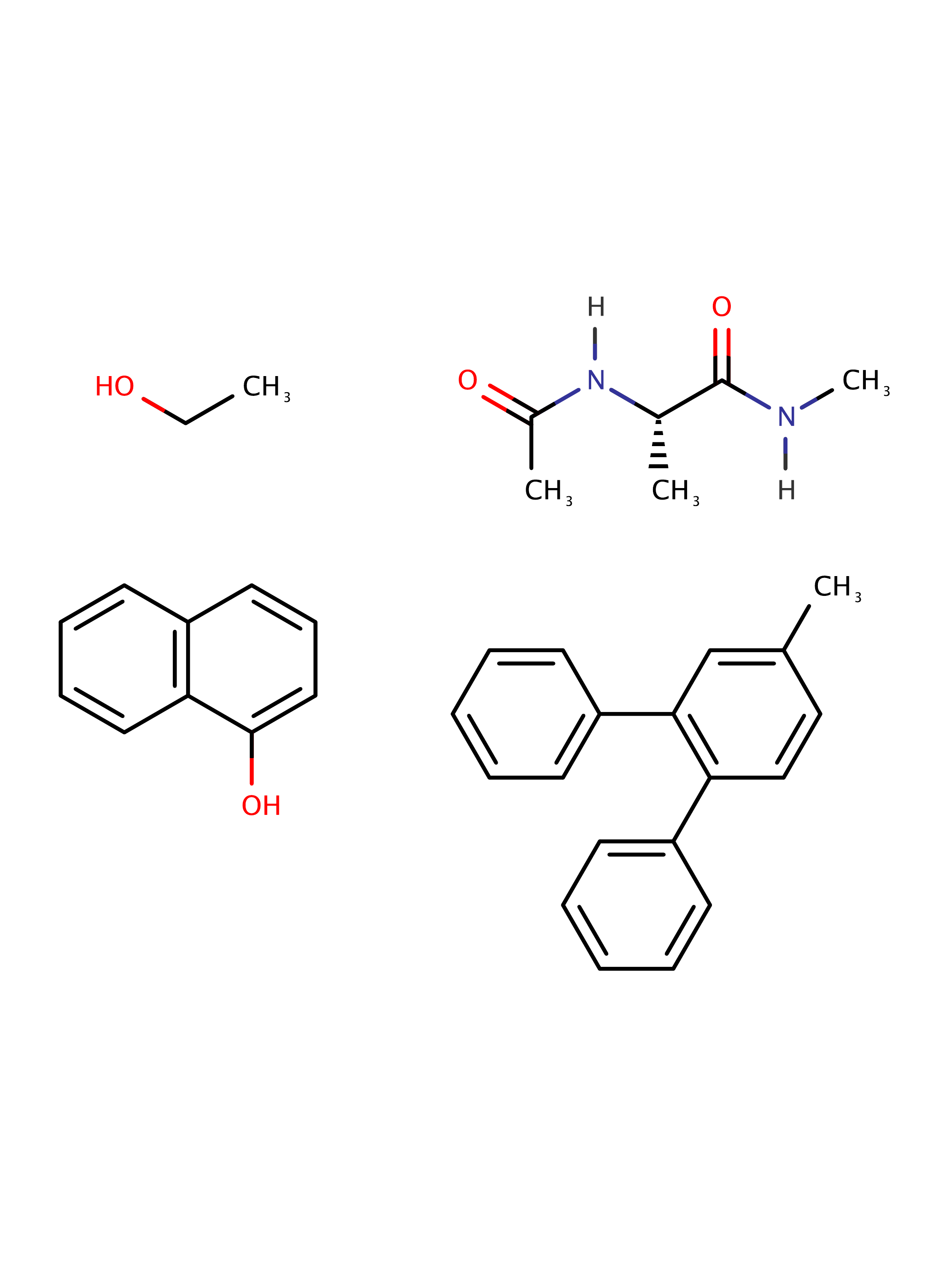}
    \caption{Chemical structures of (clockwise starting from the upper left) ethanol, alanine dipeptide, 1-naphthol, and 3,4-diphenyltoluene. The series of solutes considered is ordered here and elsewhere by increasing bulkiness and hydrophobicity.}
    \label{fig:structures}
\end{figure}

\section{\label{sec:results} Results}

\subsection{Probability Distributions of the Solute-Solvent Interaction Energy with the Linear Alchemical Potential}

\begin{figure*}
  \begin{center}
  \includegraphics{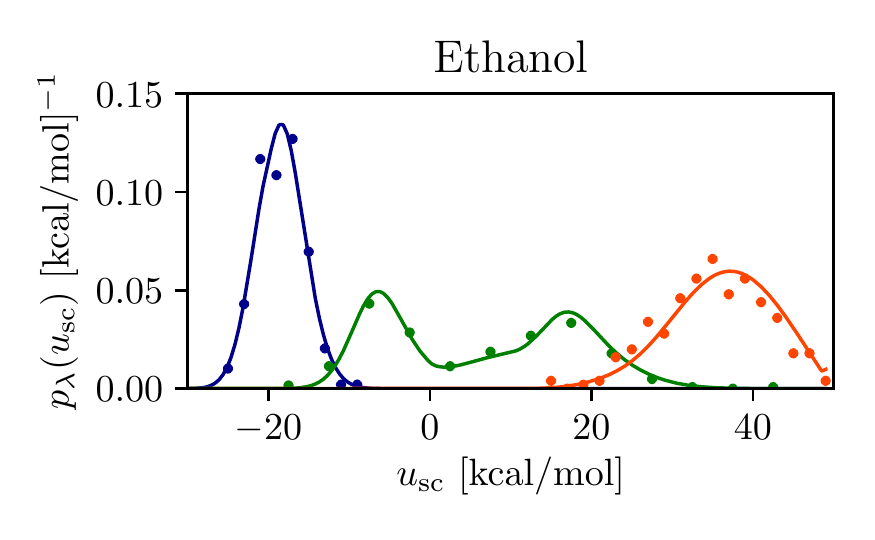}
  \includegraphics{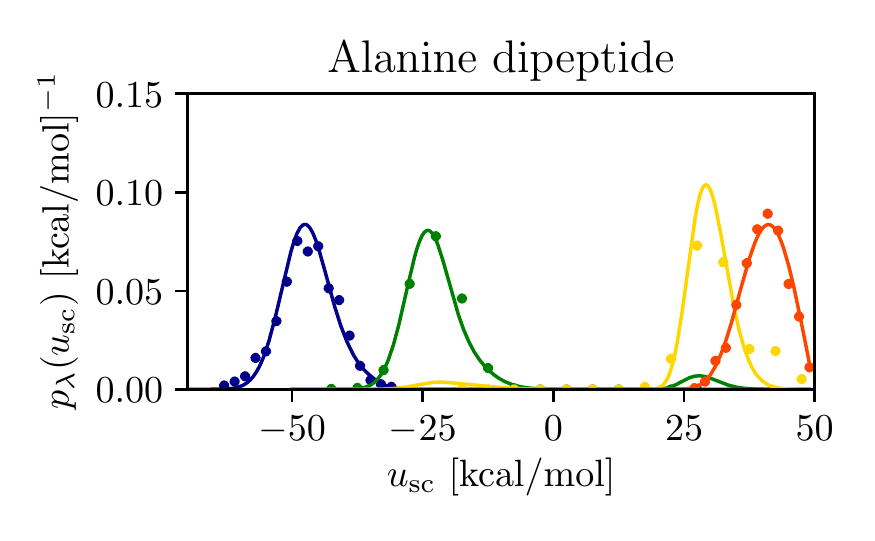}
  \includegraphics{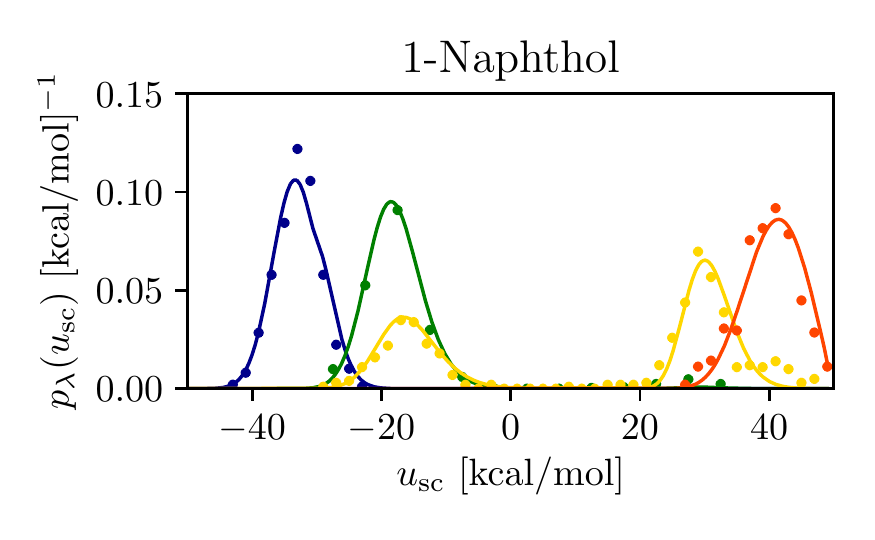}
  \includegraphics{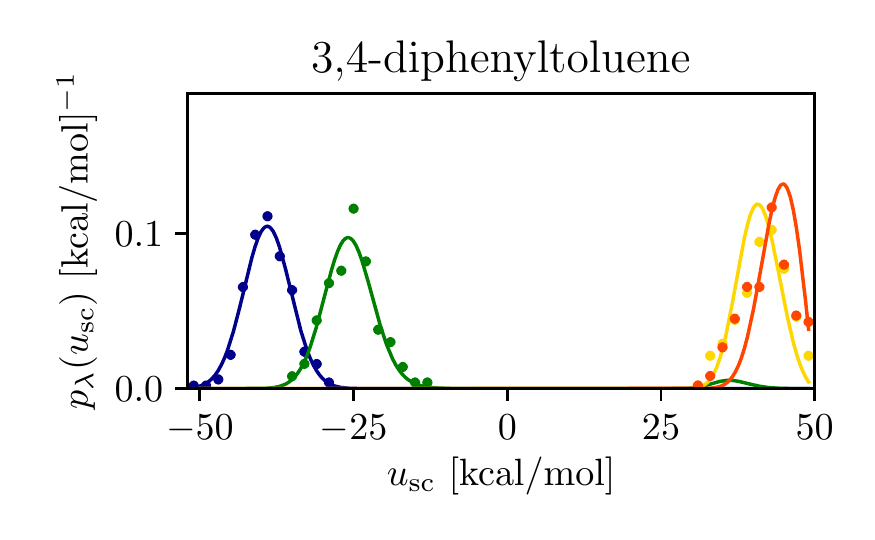}
  \end{center}

  \caption{The predicted (continuous lines) and observed (dots) probability densities of the soft-core solute-solvent interaction energy, $p_\lambda (u_{\rm sc})$, for the concerted alchemical hydration calculations with the linear alchemical potential  $W_\lambda (u_{\rm sc}) = \lambda u_{\rm sc}$ for (clockwise starting from the upper left) ethanol, alanine dipeptide, 1-naphthol, and 3,4-diphenyltoluene. Ethanol: $\lambda = 0$ (orange/red), $\lambda = 0.267$ (green), and $\lambda =1$ (blue). Alanine dipeptide: $\lambda = 0$ (orange/red), $\lambda = 0.333$ (gold), $\lambda = 0.400$ (green), and $\lambda = 1$ (blue).  1-naphthol: $\lambda = 0$ (orange/red), $\lambda = 0.333$ (gold),  $\lambda = 0.4$ (green), and $\lambda = 1$ (blue). 3,4-diphenyltoluene: $\lambda = 0$ (orange/red), $\lambda = 0.2$ (gold),  $\lambda = 0.467$ (green), and $\lambda = 1$ (blue).The predicted distributions are obtained using the analytical model for $p_0 (u_{\rm sc})$, Eqs.~(\ref{eq:superp}) to (\ref{eq:pwcaf})], with the parameters listed in Table \ref{tab:parameters}.
    \label{fig:plambda-linear}
  }
\end{figure*}

As discussed in Theory and Methods, the free energy profile is determined by the probability distributions of the perturbation energy along the alchemical path. The probability distributions $p_\lambda(u_{\rm sc})$ for selected values of $\lambda$ obtained from the concerted alchemical hydration with the linear alchemical potential are shown in Figure \ref{fig:plambda-linear} (dots). At small $\lambda$ values, the distributions are centered around large and unfavorable values of the solute-solvent interaction energy reflecting that atomic clashes are frequent when the solute and solvent are weakly coupled. Conversely, solute and solvent are more strongly coupled, and the  solute-solvent interaction energies tend to be favorable at values of $\lambda$ near 1. However, the distributions do not uniformly move towards lower values of $u_{\rm sc}$ as $\lambda$ increases. Instead, they become bimodal at intermediate values of $\lambda$ with a trough near $u_{\rm sc} = 0$ kcal/mol separating the weakly decoupled and strongly coupled states. The conversion from weakly coupled to strongly coupled behavior occurs by the transfer of population between these two limiting states rather than by the formation of  states with intermediate values of the interaction energy. This behavior is the opposite expected for linear response behavior.\cite{simonson2002gaussian} Rather, it is the hallmark of the occurrence of a strong phase transition.\cite{kim2010generalized,pal2019perturbation} 

During the concerted transformation, the transition to hydrated states occurs fairly gradually in the case of ethanol, the smallest solute considered, for which the weakly coupled and strongly coupled modes partially overlap at $\lambda = 0.4$ (Figure \ref{fig:plambda-linear}, ethanol, green dots).
However, the transition occurs sharply for all of the other solutes. For 1-naphthol, for example, the system transitions from being weakly coupled to strongly coupled in the small span of $\lambda$ from 0.333 to 0.4 (\ref{fig:plambda-linear}, 1-naphthol, yellow and green dots). Similar biphasic behavior occurs for alanine dipeptide and especially for 3,4-diphenyltoluene. Evidently, there is a critical $\lambda$ value for each solute when the weakly coupled and strongly coupled modes are equally probable.  However, it would be very difficult to pinpoint the equilibrium value accurately because interconversions from one state to the other are extremely rare (see Figure \ref{fig:bindetraj}). Indeed, the equilibration analysis described later (Table \ref{tab:transitions} and Figure \ref{fig:equilibration}) shows that, except for ethanol and perhaps alanine dipeptide, the sequence of distributions in Figure \ref{fig:plambda-linear} obtained with the linear alchemical potential is likely not converged. 

The small number of transitions between uncoupled (dehydrated) and coupled (hydrated) states is due to the small probability of visiting the so-called no man's land of interaction-energies of low probability between the two when using the concerted alchemical transformation. For 3,4-diphenyltoluene, in particular,  the probability for observing a configuration with, for example, zero solute-solvent interaction energy is immeasurably small (Figure \ref{fig:plambda-linear}, 3,4-diphenyltoluene). For the less extreme examples of alanine dipeptide and 1-naphthol, it is possible to observe distributions with mixtures of weakly coupled and strongly coupled states with very small but yet observable density in the no man's land region. However, even for these solutes, hydration and dehydration events are rarely observed in the simulations with the linear alchemical potential (Table \ref{tab:transitions} and Figure \ref{fig:bindetraj}).

Due to the lack of a sufficient number of transitions, it is not feasible to accurately estimate the relative equilibrium populations of hydrated and dehydrated states of the bulky solutes at any of the $\lambda$-states. In turn, because the hydration free energy depends on the relative populations of coupled and uncoupled states as a function of $\lambda$,\cite{kilburg2018analytical} it is expected (and largely confirmed, see below) that with the linear alchemical potential the hydration free energy estimates for the solutes except ethanol are likely to be substantially biased by finite sampling.

Collectively, this data shows that conventional linear alchemical interpolation schemes are not generally suitable for the implementation of a reliable concerted alchemical protocol. 


\subsection{Probability Distributions of the Solute-Solvent Interaction Energy with the Softplus Alchemical Potential}

The data shown in Figure \ref{fig:plambda-linear} confirms the observed probability distributions of the soft-core solute-solvent interaction energies obtained from the concerted alchemical calculations are accurately reproduced by the analytical model for $p_0(u_{\rm sc})$\cite{kilburg2018analytical,pal2019perturbation} parameterized to each dataset. The maximum likelihood-optimized parameters of the model are listed in Table \ref{tab:parameters}. The model reproduces the distributions' positions and their variations as a function of $\lambda$, including the transition points from dehydrated to hydrated states. 

 
\begin{figure*}
  \begin{center}
  \includegraphics{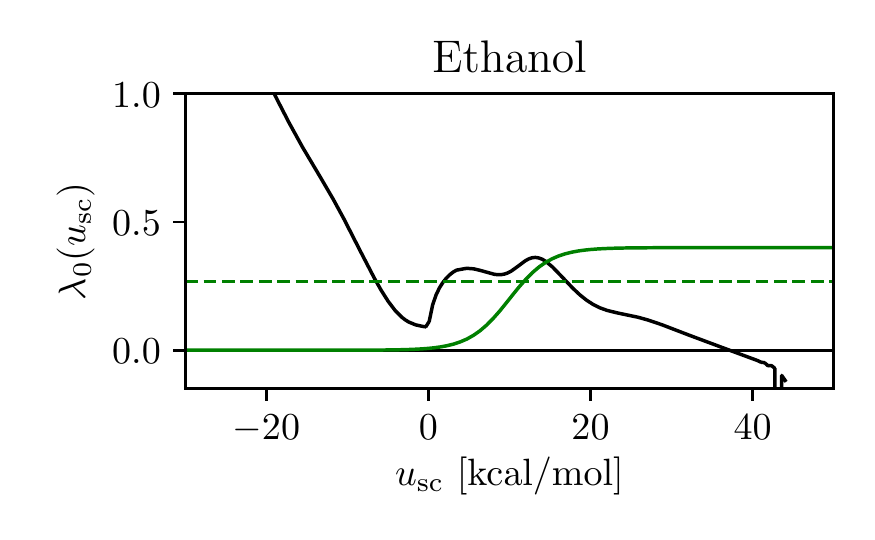}
  \includegraphics{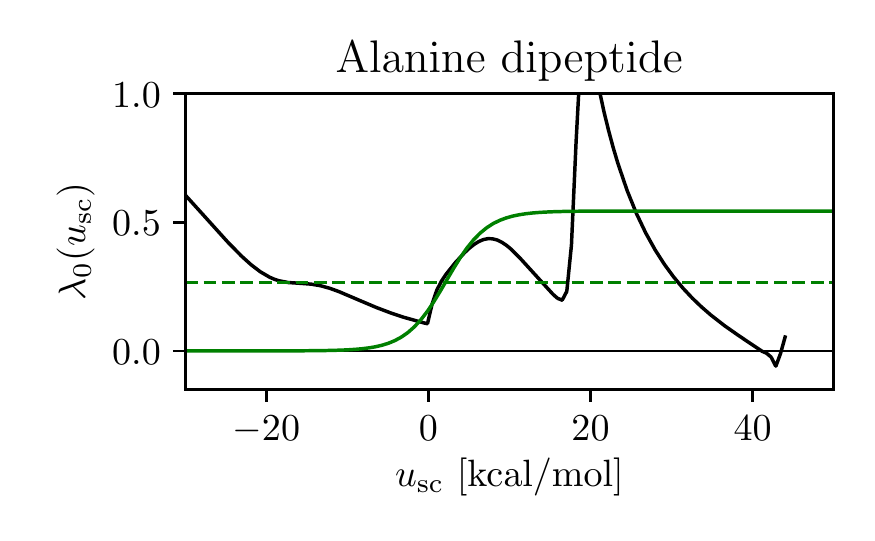}
  \includegraphics{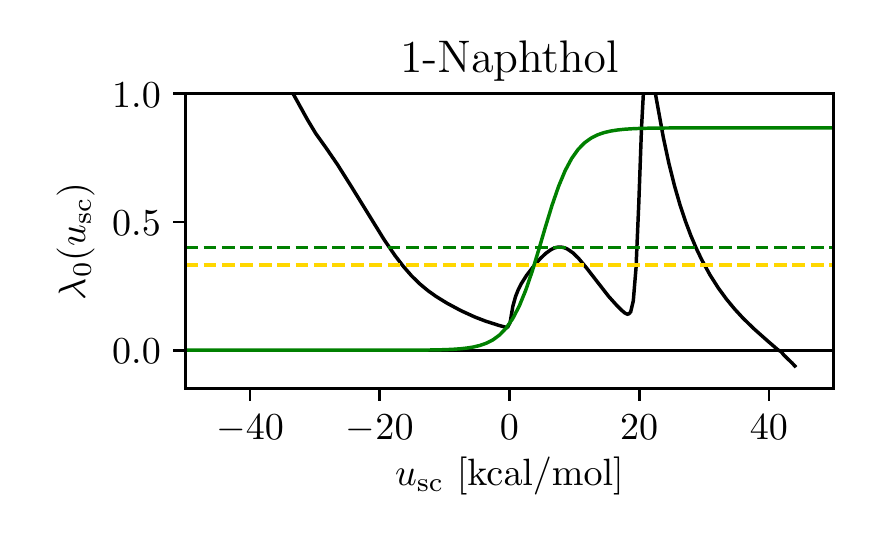}
  \includegraphics{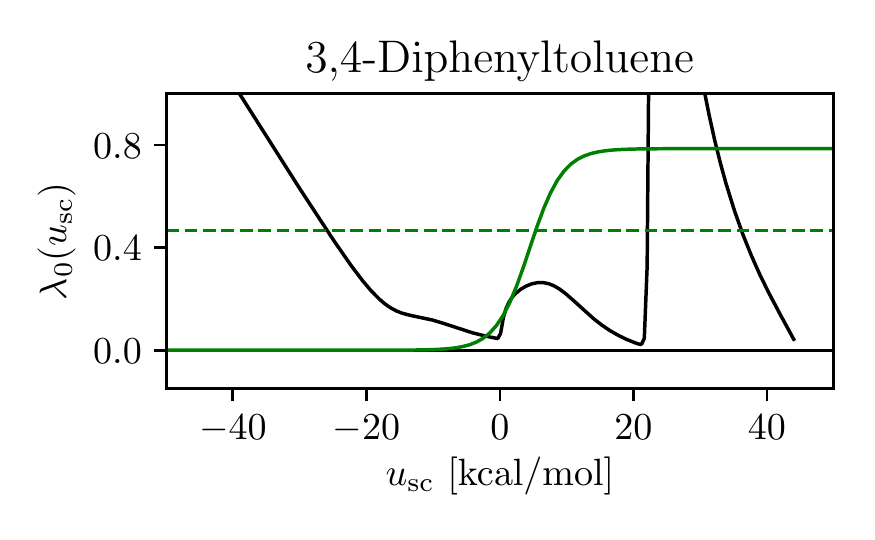}
  \end{center}

  \caption{The predicted $\lambda$-functions for the concerted hydration of the solutes indicated obtained from the analytical model for $p_0 (u_{\rm sc})$ [Eqs.~(\ref{eq:superp}) to (\ref{eq:pwcaf})], and Eq.\ (\ref{eq:lambda-function}) with the parameters listed in Table \ref{tab:parameters}. The dashed horizontal lines correspond to the values of $\lambda$ where the probability distributions of the soft-core solute solvent energy with the linear alchemical potential are observed to bimodal. Ethanol:  $\lambda = 0.267$; alanine dipeptide: $\lambda = 0.0$ and $\lambda = 0.544$, ; 1-naphthol: $\lambda = 0.333$ and $\lambda = 0.4$, 3,4-diphenytoluene: $\lambda = 0.467$. The intersections of the horizontal lines with the $\lambda$-function correspond to the maxima and minima of the respective distributions (Figure \ref{fig:plambda-linear}).\cite{pal2019perturbation} The solid sigmoid curve is the logistic function resulting from the derivative,  $\partial W_\lambda(u_{\rm sc})/\partial u_{\rm sc}$, of the softplus function at the following values of $\lambda$. Ethanol:  $\lambda = 0.267$; alanine dipeptide: $\lambda = 0.267$; 1-naphthol: $\lambda = 0.4$; 3,4-diphenyltoluene: $\lambda = 0.4$. See Tables \ref{tab:ethanol-ilog-schedule}--\ref{tab:diphenyltoluene-ilog-schedule}.
    \label{fig:lambdaf}
  }
\end{figure*}

The parameterized analytical functions for $p_0(u_{\rm sc})$ are analytically differentiated with respect to $u_{\rm sc}$ to obtain the corresponding $\lambda$-functions $\lambda_0(u_{\rm sc})$ [Eq.\ (\ref{eq:lambda-function})]\cite{pal2019perturbation} shown in Figure \ref{fig:lambdaf}. These functions are used to predict, graphically, the location of the maxima and minima of the probability distributions of the soft-core solute-solvent interaction energy as $\lambda$ is varied. The procedure consists of finding the intersections between the $\lambda$-function and, in the case of a linear perturbation potential, the horizontal line drawn at the level of the desired value of $\lambda$ [see Eq.\ (\ref{eq:lambda-function-intersection})].

Indeed, the predictions from the $\lambda$-functions in Figure \ref{fig:lambdaf} are quantitatively consistent with the observations in Figure \ref{fig:plambda-linear}. In each case, near $\lambda = 0$ the distributions are expected to have one mode at large values of $u_{\rm sc}$. As $\lambda$ is increased (that is, as the horizontal line shifts up), a back-bending region\cite{kim2010generalized} of the $\lambda$-function is encountered in which, typically, three intersections occur. The first and the last corresponding to maxima of the distributions and the middle one to a minimum. More complex patterns can arise when there are multiple back-bending regions. In the case of ethanol, the distributions are predicted to be unimodal at favorable values of the solute-solvent interaction energy above approximately $\lambda = 0.4$ (Figure \ref{fig:lambdaf}, ethanol). For all the other solutes, the back-bending of the $\lambda$-function is predicted to be so extreme to prevent, in principle, unimodal distributions with the linear potential function even for values of $\lambda$ close to $1$. While the $\lambda$-functions can be used to identify the locations of maxima and minima of the distributions, they alone do not predict the relative populations of competing modes. Thus, for example, 1-naphthol at $\lambda = 1$ is overwhelmingly more likely to form favorable interactions with the solvent even though a second, immeasurably small, mode is predicted to exist at unfavorable interaction energies.

  
\begin{figure*}
  \begin{center}
  \includegraphics{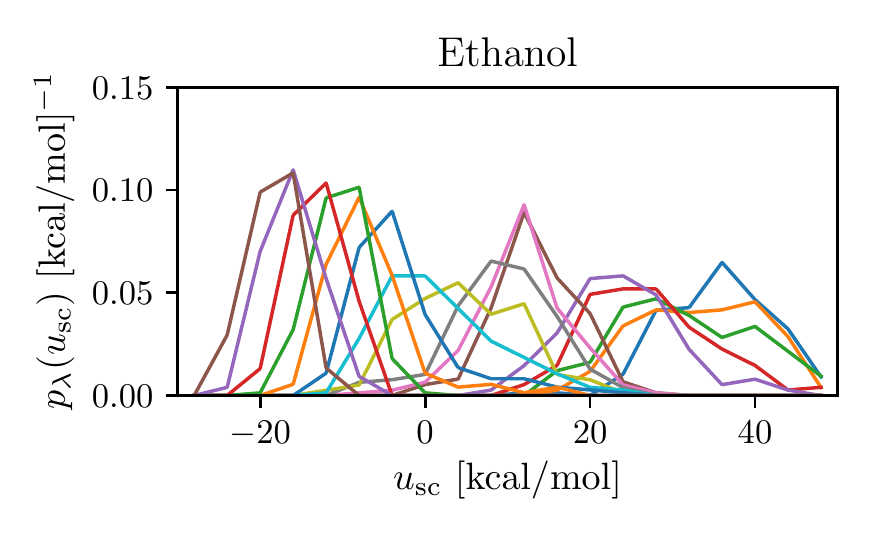}
  \includegraphics{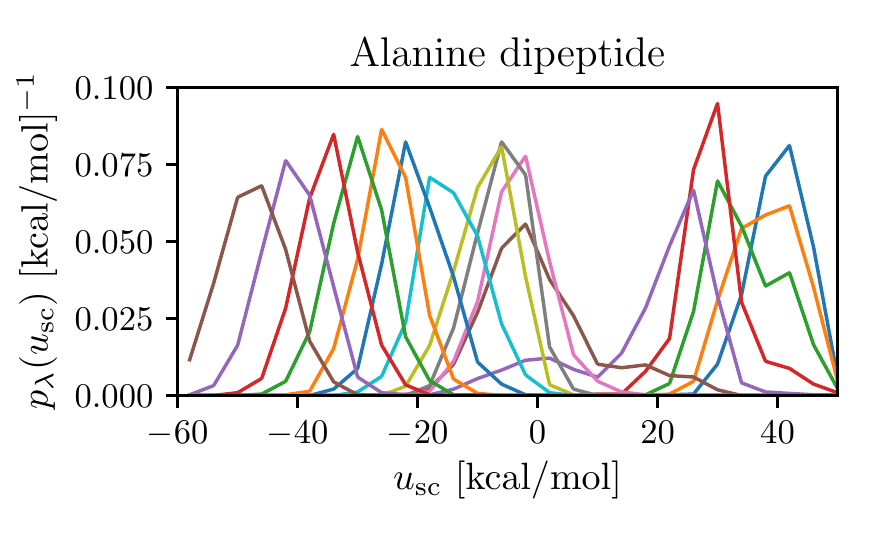}
  \includegraphics{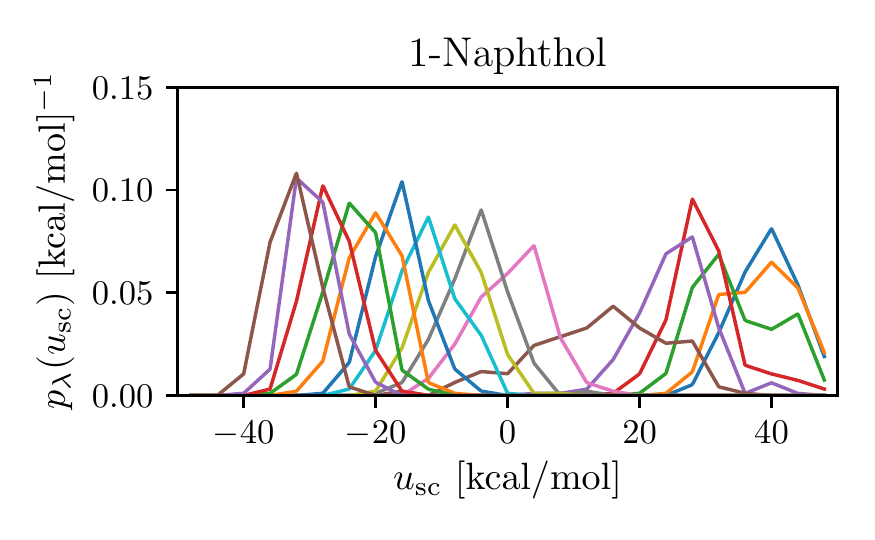}
  \includegraphics{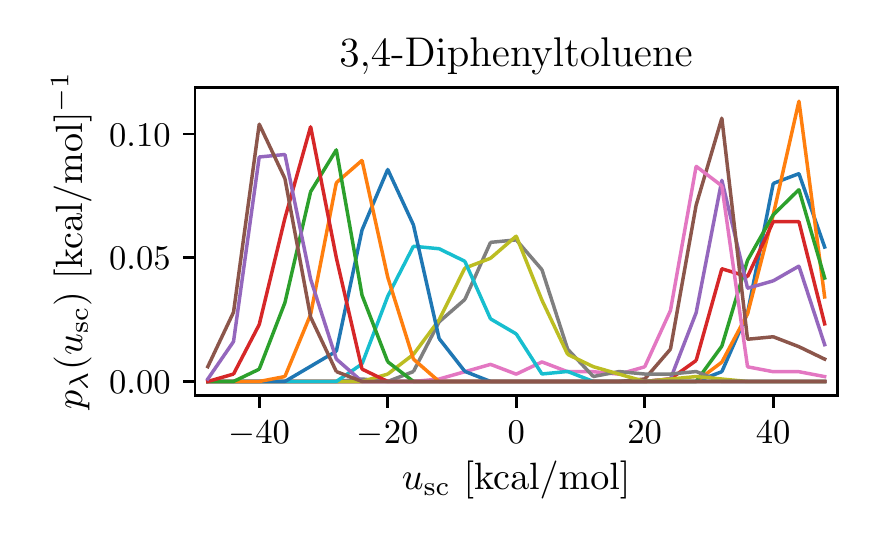}
  \end{center}

  \caption{The probability densities of the soft-core solute-solvent interaction energy obtained from the concerted alchemical hydration simulations with the softplus alchemical potential [Eq.\ (\ref{eq:ilog-function})] with the parameters in Table \ref{tab:parameters} for the hydration of the solutes indicated. The probability densities, drawn with alternating colors, shift from right to left as the $\lambda$ alchemical parameter progressively varies from $0$ (vacuum state) to $1$ (hydrated state). The probability densities shown here have been obtained from the second half of the simulation trajectories.
    \label{fig:plambda-ilog}
  }
\end{figure*}

The softplus perturbation potential can be tuned to reduce the perturbation energy gap across the hydration/dehydration transition and avoid hard-to-converge bimodal distributions.\cite{pal2019perturbation} As shown in Figure \ref{fig:lambdaf} for ethanol, the parameters of the softplus potential can be tuned to avoid multiple intersections with the $\lambda$-function, thereby avoiding the occurrence of bimodal distributions (see below). When this is not feasible, as for alanine dipeptide, naphthol, and 3,4-diphenyltoluene in  Figure \ref{fig:lambdaf}, the softplus potential can still be tuned to reduce the energy gap between competing modes and to favor one mode over another.

The alchemical simulations with the softplus potential, with the optimized parameters (Table \ref{tab:parameters}) that were designed based on the predicted $\lambda$-functions (Figure \ref{fig:lambdaf}), result in the elimination or at least the reduction of solute-solvent interaction energy gaps along the alchemical pathway. Unlike the concerted simulations with the linear alchemical potential (Figure \ref{fig:plambda-linear}), simulations with the softplus alchemical potential yield a gradual shift of the probability distributions of the solute-solvent interaction energy as the solute is coupled to the solvent (Figure \ref{fig:plambda-ilog}). The distributions with the softplus potential are also generally unimodal, which reflect either lack of multiple stable conformational macrostates or the gradual shift from one macrostate to another as $\lambda$ is varied. The softplus potential effect is not as significant for ethanol, which does not exhibit a sharp hydration transition with the linear alchemical potential (Figure \ref{fig:plambda-linear}). However, in the case of, for example, 1-naphthol, the softplus potential has a very substantial effect. In this case, an interaction energy gap of more than $30$ kcal/mol (Figure \ref{fig:plambda-linear}) is virtually eliminated (Figure \ref{fig:plambda-ilog}), thereby facilitating transitions between the weakly coupled and strongly coupled states on opposite sides of the gap (see below). As illustrated below, the hydration calculations for alanine dipeptide and 3,4-diphenyltoluene also significantly benefit from using the softplus alchemical potential, even though the no man's land with respect to the solute-solvent interaction energy is not eliminated (see Figure \ref{fig:plambda-ilog}) in these cases. 

\subsection{Analysis of Replica Exchange Efficiency}

\begin{figure*}
  \centering
  \title{\parbox{\linewidth}{\centering Linear Perturbation                Softplus Perturbation}}
  \centering
  {\Large Ethanol}
  \includegraphics[trim={1 1 1 1}]{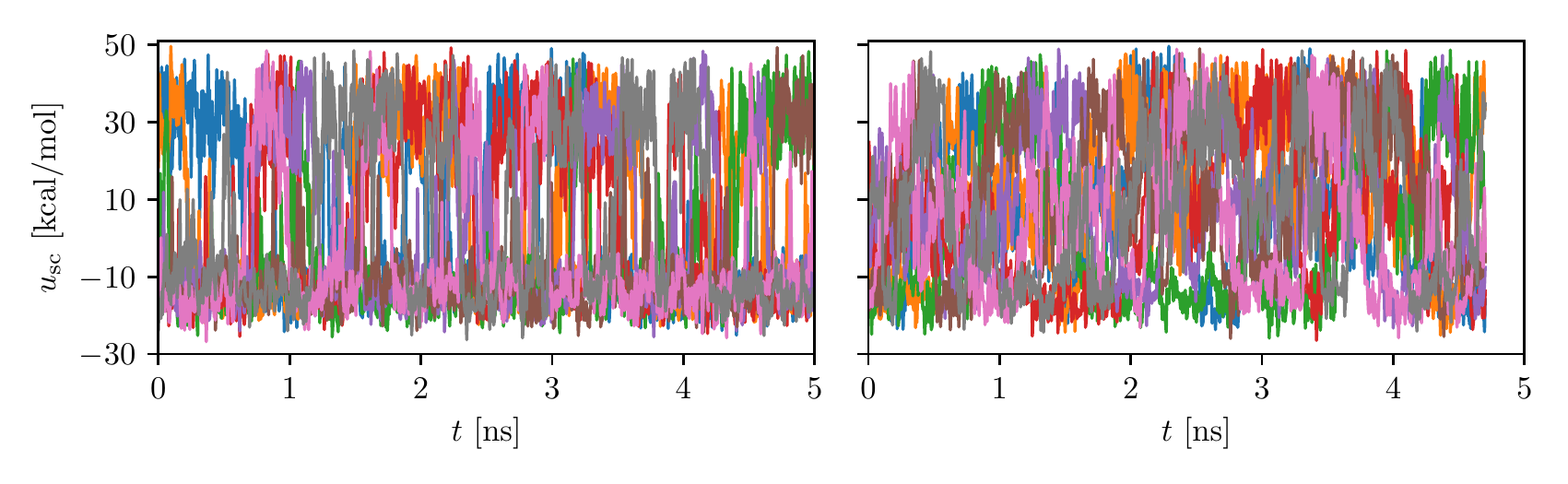}
   {\Large Alanine dipeptide}
  \includegraphics[trim={1 1 1 1}]{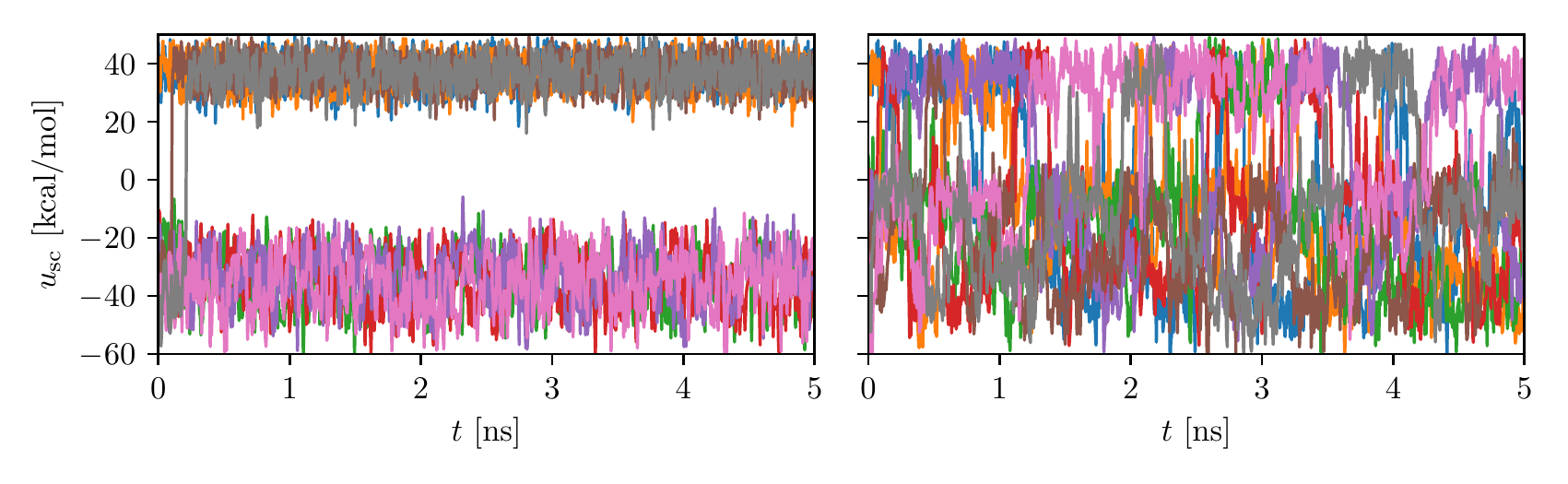}
  {\Large 1-Naphthol}
  \includegraphics[trim={1 1 1 1}]{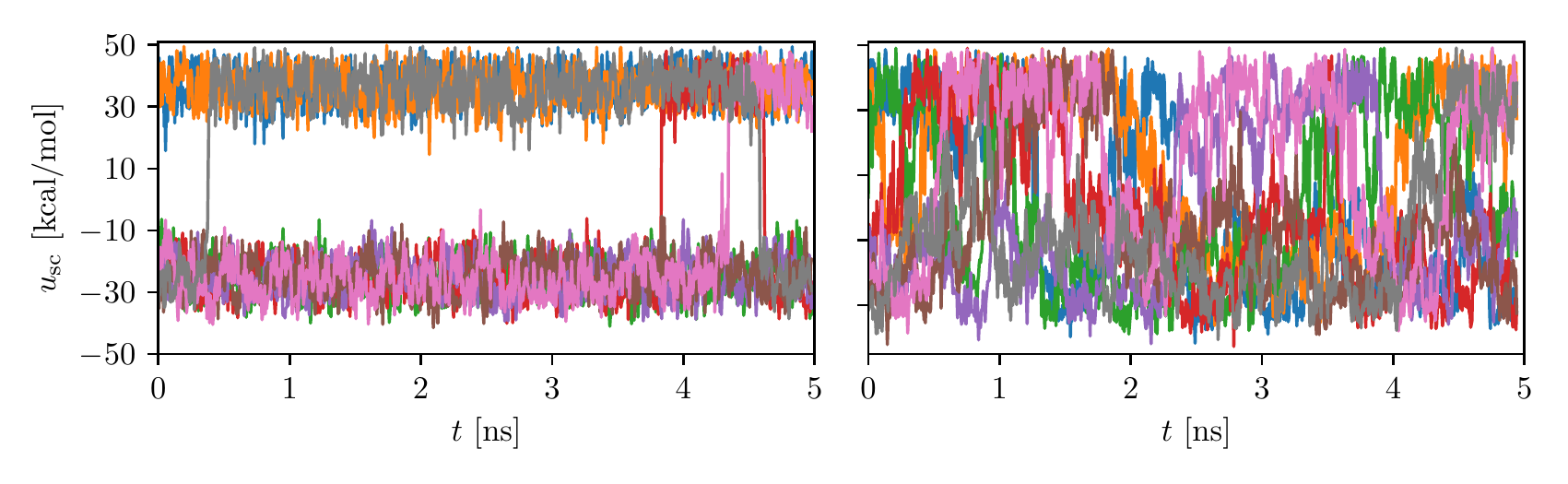}
  {\Large 3,4-Diphenyltoluene}
  \includegraphics[trim={1 1 1 1}]{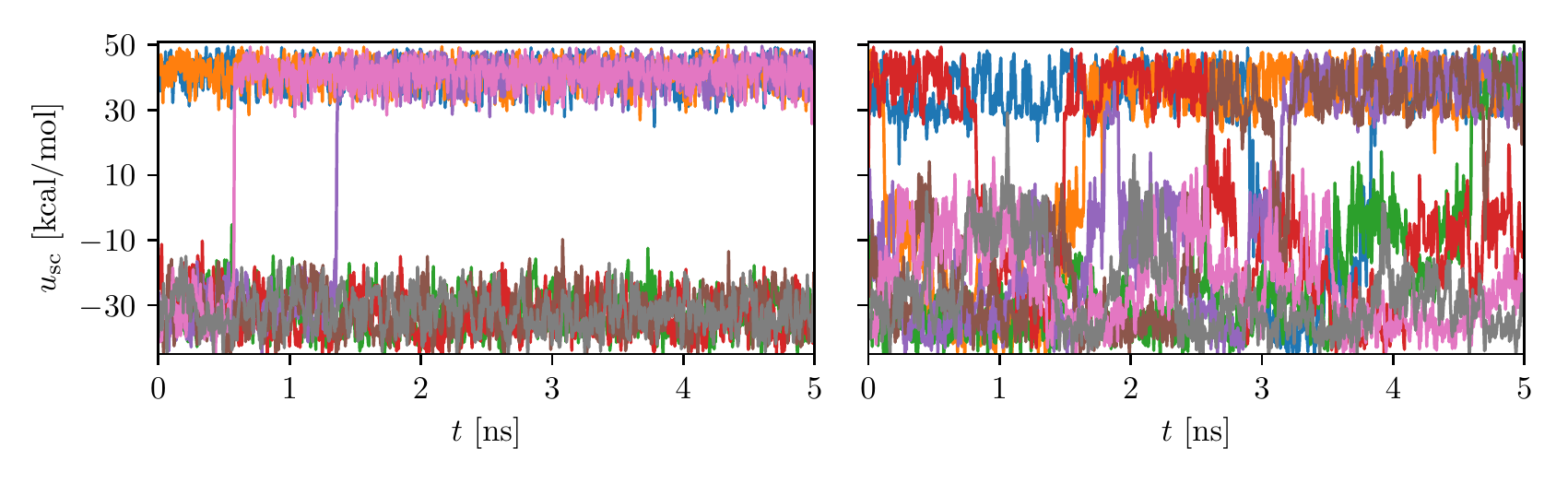}
  \caption{Time trajectories of the soft-core solute-solvent interaction energy for selected replicas of the replica exchange simulations of the alchemical hydration of ethanol, alanine dipeptide, 1-naphthol, and 3,4-diphenyltoluene with the linear (left) and softplus (right) alchemical perturbation potentials. Ethanol undergoes frequent hydration and dehydration transitions with either alchemical perturbation function. In contrast, significant numbers of hydration and dehydration transitions for the bulkier solutes (alanine dipeptide, 1-naphthol and 3,4-diphenyltoluene) are observed only with the softplus perturbation potential. 
    \label{fig:bindetraj}
  }
\end{figure*}

Hamiltonian replica exchange efficiency has been monitored here in terms of the extent of replicas' diffusion in the solute-solvent interaction energy space. In particular, we monitored the rate of transitions between coupled states with favorable solute-solvent interaction energies and uncoupled states with large and unfavorable interaction energy. The time trajectories of the soft-core solute-solvent interaction energies sampled by the replicas are shown in Figure \ref{fig:bindetraj} for each alchemical hydration simulation. The number of transitions from uncoupled to coupled states (hydration) and vice versa (dehydration) is presented in Table \ref{tab:transitions}.

The data indicate that ethanol undergoes frequent hydration/dehydration transitions with either the linear and softplus perturbation potentials. However, the bulkier solutes (alanine dipeptide, 1-naphthol, and 3,4-diphenyltoluene) undergo very few hydration/dehydration transitions with the linear alchemical potential. This behavior is due to the wide interaction energy gap between the distributions of the solute-solvent interaction energies for the the coupled and uncoupled states  (Figure \ref{fig:plambda-linear}). These solutes' bulky nature makes it very improbable for a solute molecule to find a suitable configuration in the solvent to make favorable interactions with the solvent when nearly decoupled from it. Conversely, these solutes are unlikely to overcome the energetic penalty necessary to become decoupled from the solvent when they interact with it. The coupled and decoupled states of the bulky solutes are separated by, essentially, a first-order phase transition that is entropically frustrated in the hydration direction and energetically frustrated in the dehydration direction.\cite{kim2010generalized}

\begin{table}
\caption{\label{tab:transitions} Hydration free energy estimates, equilibration times and number of hydration and dehydration transitions  for the solutes considered in this work}
\begin{ruledtabular}
\begin{tabular}{lcccc}
  Protocol & $\Delta G^\circ_h$\footnote{kcal/mol} & $t_\text{eq}$\footnote{ns} & $n_\text{hydr}$ & $n_\text{dehydr}$ \tabularnewline
\multicolumn{5}{c}{Ethanol\footnote{$u_\text{lower} = -10$, $u_\text{upper} = 25$ kcal/mol}}\tabularnewline
\multicolumn{5}{c}{Ethanol\footnote{$u_\text{lower} = -10$, $u_\text{upper} = 25$ kcal/mol}}\tabularnewline
Linear      & $-2.23 \pm 0.09$ & $0.00$  &$96$        & $96$  \tabularnewline
Softplus    & $-2.12 \pm 0.11$ & $0.63$  &$322$        & $292$  \tabularnewline
\multicolumn{5}{c}{Alanine dipeptide\footnote{$u_\text{lower} = -35$, $u_\text{upper} = 25$ kcal/mol}}\tabularnewline
Linear      & $-8.06 \pm 0.14$& $0.13$  &$0$         & $2$  \tabularnewline
Softplus    & $-8.25 \pm 0.21$& $0.00$  &$45$        & $43$  \tabularnewline
\multicolumn{5}{c}{1-Naphthol\footnote{$u_\text{lower} = -20$, $u_\text{upper} = 25$ kcal/mol}}\tabularnewline
Linear      & $-4.53 \pm 0.18$& $3.50$   &$6$         & $7$  \tabularnewline
Softplus    & $-3.66 \pm 0.14$& $0.00$   &$34$        & $32$  \tabularnewline
\multicolumn{5}{c}{3,4-Diphenyltoluene\footnote{$u_\text{lower} = -30$, $u_\text{upper} = 30$ kcal/mol}}\tabularnewline
Linear      & $-0.33 \pm 0.18$& $1.38$  & $0$         & $2$  \tabularnewline
Softplus    & \hspace{2.5 mm}$3.51 \pm 0.26$& $3.38$  & $15$        & $14$  \tabularnewline

\end{tabular} 
\end{ruledtabular}  
\end{table}

On the other hand, the optimized softplus perturbation potential is very effective at circumventing the phase transition. As shown in Figure \ref{fig:bindetraj} and Table \ref{tab:transitions}, many more hydration/dehydration transitions are observed with the softplus potential than with the linear potential. Many transitions make it possible to equilibrate and converge the hydration free energy estimates for the bulkier solutes more rapidly without additional computational expense (see below).

\subsection{Equilibration of the Hydration Free Energy Estimates}

\begin{figure*}
  \centering

  {\Large Ethanol}
  \includegraphics{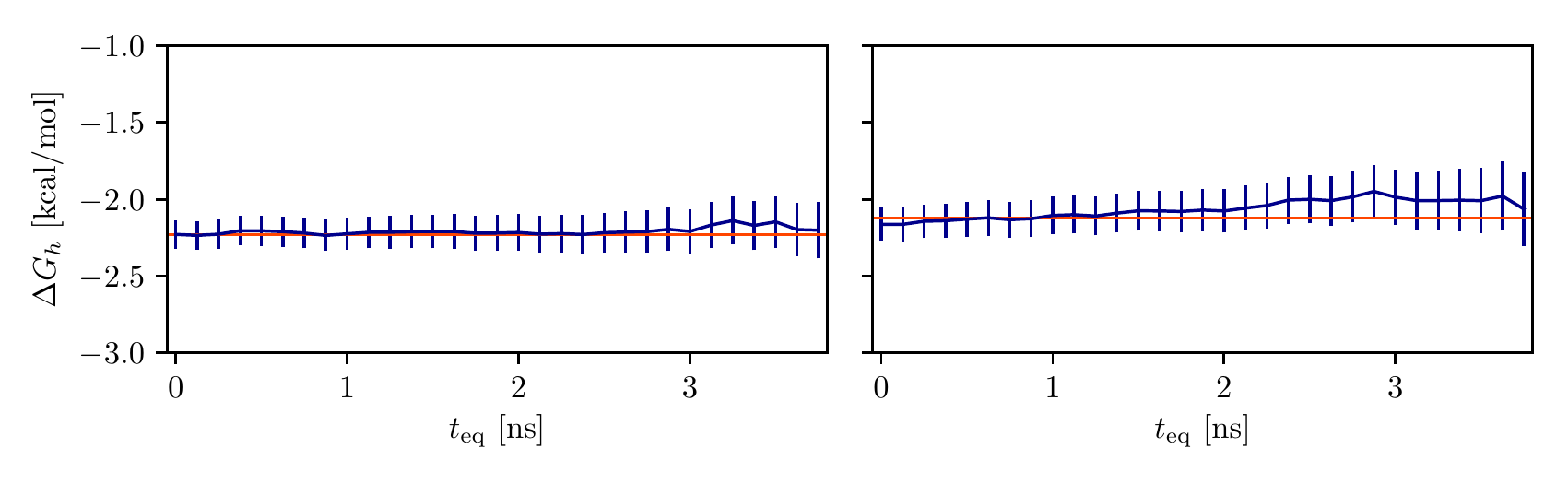}
  {\Large Alanine dipeptide}
  \includegraphics{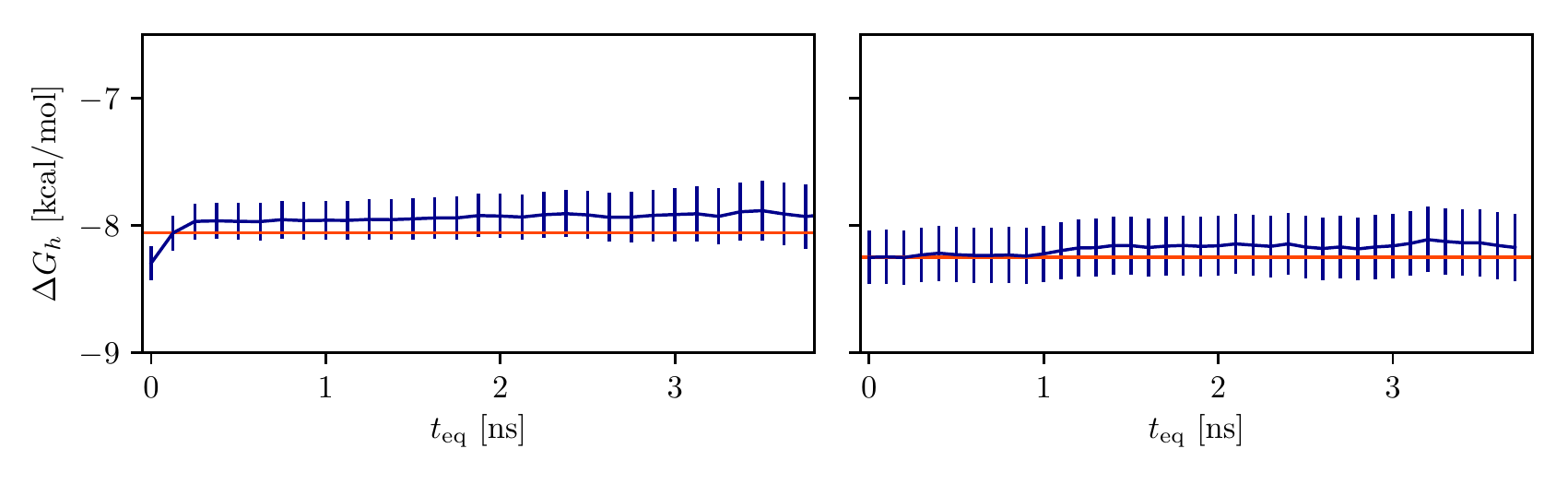}
  {\Large 1-Naphthol}
  \includegraphics{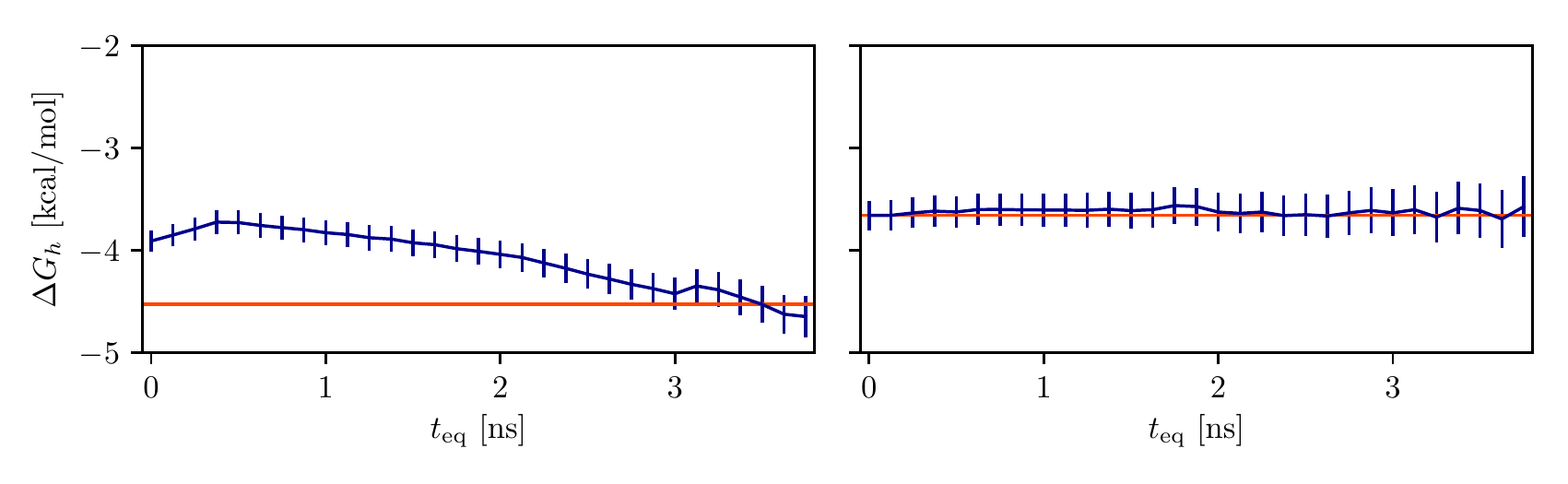}
  {\Large 3,4-Diphenyltoluene}
  \includegraphics{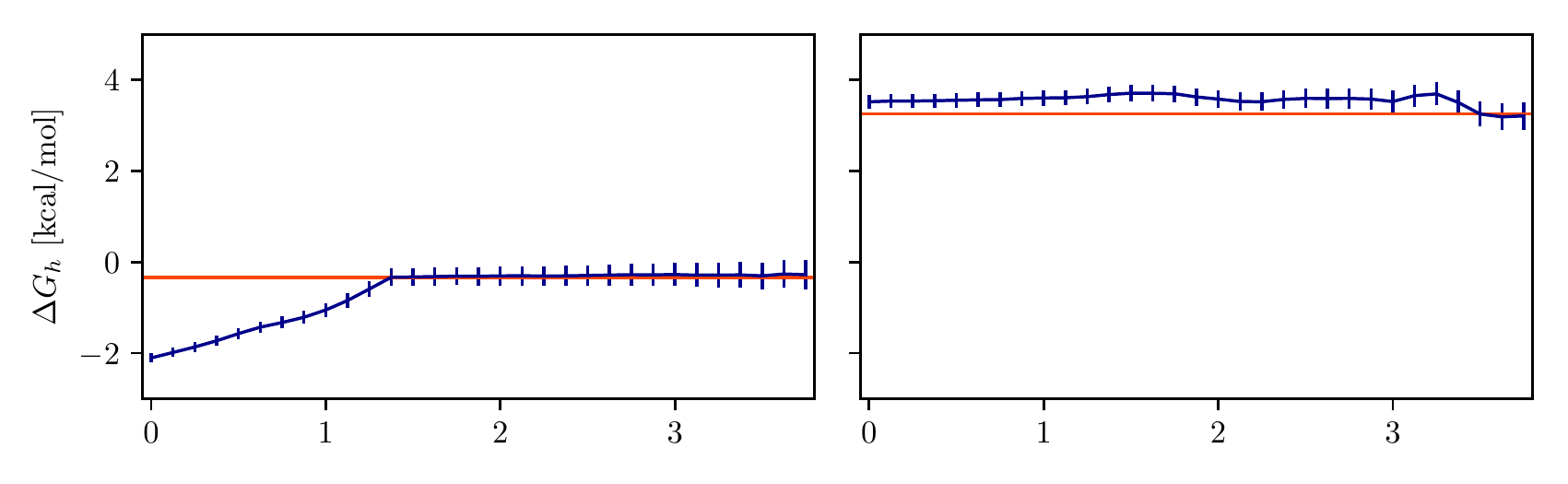}
  \caption{Reverse cumulative equilibration profiles of the hydration free energies of the solutes indicated with the linear (left) and softplus (right) alchemical perturbation potentials. The horizontal line corresponds to the value of the estimate of the hydration free energy for each case, defined as the hydration free energy at the earliest equilibration time corresponding to an estimate  statistically consistent with those of all of the following equilibration times. 
    \label{fig:equilibration}
  }
\end{figure*}

The computed hydration free energy estimates are plotted in Figure \ref{fig:equilibration} as a function of the amount of data discarded from the start of the simulation (the equilibration time). These plots, referred to as reverse cumulative equilibration profiles,\cite{yang2004free,chodera2016simple,kilburg2018assessment} are used to determine the time after which the time series of data generated by the simulation becomes stationary and not biased by the starting conformation of the system. It is not obvious to pinpoint such a time because, while the accuracy of the binding free energy presumably improves as more non-equilibrated samples are discarded, the precision of the estimate (indicated by the error bars in Figure \ref{fig:equilibration}) worsens as more initial samples are discarded. Here we take the approach of choosing the hydration free energy estimate as the one corresponding to the shortest equilibration time that gives a value statistically indistinguishable from those at all subsequent equilibration times.\cite{kilburg2018assessment}

Based on this criterion, we conclude that the hydration free energy estimate for ethanol equilibrates almost immediately after the start of the simulation with either the linear or the softplus alchemical potentials. The hydration free energy estimates for ethanol with the linear and softplus potentials are in statistical agreement (Table \ref{tab:transitions}). Given the very different nature of the alchemical paths in the two simulations, we conclude that equilibration and convergence of the hydration free energy have been achieved in this case. This positive outcome is consistent with the high rate of hydration and dehydration transitions observed  (Figure \ref{fig:bindetraj} and Table \ref{tab:transitions}) for ethanol with either the linear and softplus potentials.  The result for ethenol confirms the validity of the concerted alchemical protocol and the correctness of the simulation algorithms in producing a canonical ensemble of conformations with either potential. 

The same analysis shows that equilibration of the concerted free energy protocol is achieved rapidly for the bulkier solutes (alanine dipeptide, 1-naphthol, and 3.4-diphenyltoluene) when using the optimized softplus potential (Figure \ref{fig:equilibration}, second column). The hydration free energy estimates are deemed well converged also based on the relatively large hydration/dehydration transitions observed for these solutes. 3,4-diphenyltoluene, the largest and most hydrophobic solute, is confirmed as a stringent test case for one-step concerted alchemical protocols.\cite{lee2020improved} Despite the relatively few hydration/dehydration transitions observed even after optimization of the softplus potential, the approach employed here appears to yield a converged and reliable hydration free energy estimate for this compound. After taking into account long-range van der Waals interactions, which are expected to contribute as much as 3 kcal/mol of additional solute-solvent interaction energy, the hydration free energy obtained here for 3,4-diphenyltoluene in a droplet is consistent with earlier calculations with periodic boundary conditions.\cite{lee2020improved} To our knowledge, the measured hydration free energy for this compound has never been reported.

In contrast to the softplus potential, the concerted hydration alchemical protocol fails to converge for the bulkier solutes when using the linear alchemical potential. This is indicated by the very few hydration/dehydration transitions (Table \ref{tab:transitions}) and the cumulative equilibration profiles (Figure \ref{fig:equilibration}, first column, alanine dipeptide, 1-naphthol, and 3,4-diphenyltoluene). The rapid equilibration of the hydration free energy estimate for alanine dipeptide with the linear potential is likely circumstantial as it hinges on only two dehydration transitions early in the replica exchange trajectories (Figure \ref{fig:bindetraj}, alanine dipeptide, left panel). With the linear perturbation potential, the hydration free energy of 1-naphthol appears to equilibrate after $3$ ns at a value of $-4.53 \pm 0.18$ kcal/mol, which is statistically inconsistent with the less biased estimate of $-3.66 \pm 0.14$ obtained with the softplus potential (Table \ref{tab:transitions} and Figure \ref{fig:equilibration}, 1-naphthol). The concerted alchemical hydration of 3,4-diphenyltoluene is a good example of the potential convergence pitfalls of free energy calculations. With the linear potential  (Figure \ref{fig:equilibration}, 3,4-diphenyltoluene), the hydration free energy appears to equilibrate after approximately 1.3 ns at a very stable plateau value, which could be mistaken as a converged estimate. In actuality, the value of the plateau is not converged as confirmed by the lack of hydration/dehydration transition events after 1.3 ns of simulation per replica (Figure \ref{fig:bindetraj}, 3,4-diphenyltoluene, left panel).

The data collected for these solutes strongly supports the finding that commonly used linear alchemical perturbation potentials are unsuitable for obtaining converged free energy estimates with a concerted alchemical process. In contrast, optimized softplus potentials can enable rapid equilibration and convergence by promoting hydration/dehydration transitions.

\section{\label{sec:discussion} Discussion}

Alchemical hydration free energy calculations are widely employed to predict the water solubilities of substances,\cite{mobley2014freesolv,yu2014order} test force field and free energy protocols,\cite{mobley2012alchemical,Mobley2007b,shivakumar2010prediction} and to estimate absolute and relative binding free energies of molecular complexes.\cite{Gilson:Given:Bush:McCammon:97,Chodera:Mobley:cosb2011} While alternatives have been proposed,\cite{christ2008multiple,lee2020improved} the generally accepted guideline to avoid slow convergence and numerical instabilities are to split the alchemical hydration process into two steps. In the first step, volume-exclusion and dispersion solute-solvent interactions are turned on, followed by a second step in which electrostatic interactions are established.\cite{klimovich2015guidelines} In addition, especially during the volume-exclusion phase, it has been found necessary to employ custom soft-core interatomic pair potentials.\cite{pitera2002comparison,Steinbrecher2011,konig2020alternative}

These free energy practices, which are widely implemented in popular molecular simulation packages,\cite{brooks2009charmm,phillips2005scalable,pronk2013gromacs,eastman2017openmm,he2020fast} are generally successful and robust. Nevertheless, it would be beneficial to explore concerted alternatives with improved equilibration and convergence characteristics. The software implementations of free energy methods can also be cumbersome and challenging to integrate and maintain alongside other molecular simulation algorithms. A recent study by Lee et al.\cite{lee2020improved} discussed the downsides of multi-step free energy methods and called for a streamlined concerted approach that would be more easily integrated with extended ensemble and self-adjusting conformational sampling algorithms, as well as non-equilibrium approaches.\cite{darve2008adaptive,procacci2019solvation,procacci2020virtual} Lee et al.\cite{lee2020improved} proposed a family of soft-core pair potentials and non-linear alchemical hybrid perturbations that facilitates the calculation of hydration free energies in a concerted fashion. 

As shown in this work, concerted alchemical protocols present unique computational challenges that have likely discouraged progress in earlier attempts. We have shown for example that conventional linear alchemical interpolating potentials can introduce severe entropic bottlenecks that preclude the rapid equilibration between the hydrated and dehydrated states of the solute, resulting in strongly biased free energy estimates.

In a recent study,\cite{pal2019perturbation} we identified pseudo first-order phase transitions along the alchemical path as the critical and fundamental obstacles to the application of concerted alchemical processes for molecular binding. In the same study, we addressed the issue by applying Hamiltonian-shaping techniques inspired by non-Boltzmann sampling methodologies developed by Straub and collaborators\cite{kim2010generalized,lu2013order} for the study of temperature-driven first order phase transitions. Using this approach, we were able to identify alchemical paths that avoid or soften biphasic behavior in the context of binding with an implicit description of the solvent.\cite{pal2019perturbation} In the present study, we demonstrate that these same techniques can also be applied to the estimation of hydration free energies with explicit solvation. Critical to this approach are a new family of alchemical perturbation functions presented here that are of potentially broad applicability. We also illustrate a theoretical framework and a graphical procedure to optimize them to systematically improve conformational sampling and the rate of equilibration and convergence of free energy estimates. 

In this work, we considered the hydration of solutes into a water droplet.  This simplified system is a first, yet representative validation of the method. The droplet model was selected mainly for expediency to validate the theory and the model without extensive modifications of our Single Decoupling Method (SDM) software originally developed alongside an implicit description of the solvent.\cite{pal2019perturbation} In SDM, the perturbation from the coupled state to the uncoupled state of the molecular complex is achieved by displacing the ligand out of the receptor binding pocket and into the surrounding implicit solvent bulk. The same approach would not work for decoupling the solute molecule from an explicit solvent system with periodic boundary conditions since a periodic boundary system does not have an outer region where to place the solute. A software implementation of our method with periodic boundary conditions would require the implementation of solute-solvent decoupling through the tuning of force field parameters. While OpenMM supports this procedure,\cite{wang2013yank} our immediate focus is to build upon this result to develop concerted binding free energy estimation protocols with explicit solvation and periodic boundary conditions. This work is ongoing and will be reported in upcoming publications.

The theoretical, algorithmic, and numerical simulation strategies presented and the promising results illustrated here indicate a possible roadmap to streamline the implementation of alchemical protocols in molecular simulation packages. The first suggestion is to replace soft-core pair potentials with, as done here, a soft-core function applied to the solute-solvent interaction energy [e.g. using Eqs.\ (\ref{eq:soft-core-general}) and (\ref{eq:rat-sc})]. This approach would greatly simplify the implementation of alchemical protocols in molecular dynamics packages by making it unnecessary to modify the core energy and force subroutines. While the method used here requires re-processing the forces,  this is cleanly and efficiently accomplished in our implementation by a $O(N)$ loop coded in a separate small subroutine of the molecular dynamics integrator. The family of alchemical potentials introduced here also simplifies the thermodynamic reweighting procedure with MBAR or UWHAM\cite{Shirts2008a,Tan2012} for free energy estimation. Alchemical implementations with soft-core pair potentials require the re-evaluation, using the molecular dynamics engine, of each sample's perturbation energy for each $\lambda$ state on the alchemical path.\cite{li2020repulsive} This cumbersome process is completely bypassed in our implementation. The alchemical potential introduced here is a one-to-one function of the solute-solvent interaction energy, which can be saved together with each trajectory frame and manipulated algebraically with minimal effort, using Eq.\ (\ref{eq:pert_pot}), to yield the perturbation energies at any of the alchemical states.

The second suggestion emerging from this study is using a non-linear alchemical perturbation that, like the softplus potential we propose [Eq.\ \ref{eq:ilog-function}], judiciously warps the alchemical path in critical regions to avoid phase transitions while retaining a linear character away from the problematic regions. Furthermore, we suggest the use of the $\lambda$-function formalism [Eqs.\ (\ref{eq:lambda-function}) and (\ref{eq:lambda-function-intersection})] and adaptive maximum likelihood-based approaches to systematically optimize the alchemical schedule and the alchemical perturbation function to enhance the rate of equilibration and convergence.

\section{Conclusions}

We employ a statistical analytical theory of molecular association\cite{kilburg2018analytical} and a single-decoupling alchemical method for binding free energy estimation with implicit solvation\cite{pal2019perturbation,kim2010generalized} to develop a concerted alchemical protocol for the calculation of hydration free energies of small to medium-sized molecules with explicit solvation. The concerted alchemical hydration protocol introduced here involves a single alchemical transformation rather than the more commonly employed pair of distinct decoupling transformations for electrostatic and steric/dispersion interactions. The approach and its benefits are illustrated for model systems involving the hydration of four diverse solutes in a water droplet. This proof of principle study paves the way for a new generation of streamlined concerted free energy estimation algorithms in condensed phases.  

\begin{acknowledgments}

 We acknowledge support from the National Science Foundation (NSF
CAREER 1750511 to E.G.). Molecular simulations were conducted on the
Comet GPU supercomputer cluster at the San Diego Supercomputing Center
supported by NSF XSEDE award TG-MCB150001.
\end{acknowledgments}

\appendix

\section{Appendix}

\subsection{Parameters of the Analytical Model for $p_0(u_{\rm sc})$}

See Table \ref{tab:parameters}.

\begin{table*}[p]
  \caption{\label{tab:parameters}Optimized parameters for the analytical model of molecular association for the two systems studied in this work. Uncertainties are implied by the number of reported significant figures.}
\begin{ruledtabular}
\begin{tabular}{lccccccc}
       & weight               & $p_b$ & $\bar{u}_{b}$\footnote{In kcal/mol} & $\sigma_{b}^{\rm a} $ & $\epsilon^{\rm a}$ & $\tilde{u}^{\rm a}$ & $n_l$\tabularnewline
\multicolumn{8}{c}{Ethanol}\tabularnewline
mode 1 & $3.41 \times 10^{-3}$ & $1.13 \times 10^{-6}$ & $-7.30$             & $2.68$              & $5.83$           &   $-0.07$          & $4.08$ \tabularnewline
mode 2 & $2.09 \times 10^{-1}$ & $9.40 \times 10^{-7}$ & $-3.70$             & $2.70$              & $28.6$           &   $1.90$           & $4.41$ \tabularnewline
mode 3 & $6.11 \times 10^{-1}$ & $2.27 \times 10^{-9}$ & $5.83$                 & $1.02$               & $22.7$           &   $22.7$             & $9.84$  \tabularnewline
\multicolumn{8}{c}{Alanine dipeptide}\tabularnewline
mode 1 & $1.16 \times 10^{-10}$ & $5.33 \times 10^{-8}$ &$-16.32$            & $4.32$              & $4.38$           &   $1.28$           & $5.20$ \tabularnewline
mode 2 & $2.03 \times 10^{-7}$ & $1.62 \times 10^{-8}$ & $-14.31$            & $3.88$              & $3.5$           &   $10.8$          & $5.93$ \tabularnewline
mode 3 & $1.00$                & $0$                   & $103.89$               & $4.72$                & $33.6$           &   $83.0$            & $19.07$  \tabularnewline
\multicolumn{8}{c}{1-Naphthol}\tabularnewline
mode 1 & $2.92 \times 10^{-8}$ & $7.90 \times 10^{-8}$ & $-14.86$            & $3.36$              & $4.24$           &   $1.85$           & $4.85$ \tabularnewline
mode 2 & $9.92 \times 10^{-7}$ & $7.68 \times 10^{-8}$ & $-12.13$            & $3.32$              & $6.84$           &   $11.8$          & $5.06$ \tabularnewline
mode 3 & $1.00$                & $0$                   & $198$               & $3$                 & $9.50$           &   $250$            & $19.1$  \tabularnewline
\multicolumn{8}{c}{3,4-Diphenyltoluene}\tabularnewline
mode 1 & $1.06 \times 10^{-15}$ & $4.54 \times 10^{-5}$ & $-14.7$            & $3.80$              & $3.66$           &   $13.8$           & $3.14$ \tabularnewline
mode 2 & $1.00$ & $0$ & $172$            & $4.00$              & $5.82$           &   $367$          & $49.2$ \tabularnewline
\end{tabular} 
\end{ruledtabular}
\end{table*}

\subsection{Alchemical Schedule and Parameters of the Softplus Perturbation Function}

See Tables \ref{tab:ethanol-ilog-schedule} and \ref{tab:naphtol-ilog-schedule}.

\begin{table}[p]
  \caption{\label{tab:ethanol-ilog-schedule} Alchemical schedule of the softplus perturbation function for the hydration of ethanol.}
\begin{ruledtabular}
\begin{tabular}{lccccc}
  $\lambda$  &   $\lambda_1$ & $\lambda_2$ & $\alpha$\footnote{kcal/mol$^{-1}$} & $u_0$\footnote{kcal/mol} & $w_0$\footnote{kcal/mol}  \tabularnewline
 0.000 &   0.000 &   0.000 &   0.400 &  10.000 &   0.000  \tabularnewline
 0.067 &   0.000 &   0.044 &   0.400 &  10.000 &  -0.521  \tabularnewline  
 0.133 &   0.000 &   0.089 &   0.400 &  10.000 &  -1.043  \tabularnewline
 0.200 &   0.000 &   0.133 &   0.400 &  10.000 &  -1.564  \tabularnewline
 0.267 &   0.000 &   0.178 &   0.400 &  10.000 &  -2.086  \tabularnewline
 0.333 &   0.000 &   0.400 &   0.400 &   8.889 &  -4.249  \tabularnewline
 0.400 &   0.000 &   0.400 &   0.400 &   6.667 &  -3.360  \tabularnewline
 0.467 &   0.000 &   0.400 &   0.400 &   4.444 &  -2.471  \tabularnewline
 0.533 &   0.000 &   0.400 &   0.400 &   2.222 &  -1.582  \tabularnewline
 0.600 &   0.000 &   0.400 &   0.400 &   0.000 &  -0.693  \tabularnewline
 0.667 &   0.167 &   0.400 &   0.400 &   0.000 &  -0.404  \tabularnewline
 0.733 &   0.333 &   0.400 &   0.400 &   0.000 &  -0.116  \tabularnewline
 0.800 &   0.500 &   0.500 &   0.400 &   0.000 &   0.000  \tabularnewline
 0.867 &   0.667 &   0.667 &   0.400 &   0.000 &   0.000  \tabularnewline
 0.933 &   0.833 &   0.833 &   0.400 &   0.000 &   0.000  \tabularnewline
 1.000 &   1.000 &   1.000 &   0.400 &   0.000 &   0.000  \tabularnewline
\end{tabular} 
\end{ruledtabular}
\end{table}

\begin{table}[p]
  \caption{\label{tab:dialanine-ilog-schedule} Alchemical schedule of the softplus perturbation function for the hydration of alanine dipeptide.}
\begin{ruledtabular}
\begin{tabular}{lccccc}
  $\lambda$  &   $\lambda_1$ & $\lambda_2$ & $\alpha$\footnote{kcal/mol$^{-1}$} & $u_0$\footnote{kcal/mol} & $w_0$\footnote{kcal/mol}  \tabularnewline
 0.000 &  0.000 & 0.000 &  0.400 &  3.000 &  0.000  \tabularnewline
 0.067 &  0.000 & 0.138 &  0.400 &  3.000 & -0.778  \tabularnewline
 0.133 &  0.000 & 0.258 &  0.400 &  3.000 & -1.557  \tabularnewline
 0.200 &  0.000 & 0.395 &  0.400 &  3.000 & -2.335  \tabularnewline
 0.267 &  0.000 & 0.544 &  0.400 &  2.200 & -3.113  \tabularnewline
 0.333 &  0.000 & 0.651 &  0.400 &  2.200 & -3.892  \tabularnewline
 0.400 &  0.000 & 0.759 &  0.400 &  2.200 & -5.837  \tabularnewline
 0.467 &  0.000 & 0.867 &  0.400 &  1.100 & -2.947  \tabularnewline
 0.533 &  0.067 & 0.867 &  0.400 &  0.000 & -1.387  \tabularnewline
 0.600 &  0.200 & 0.867 &  0.400 &  0.000 & -1.156  \tabularnewline
 0.667 &  0.333 & 0.867 &  0.400 &  0.000 & -0.925   \tabularnewline
 0.733 &  0.467 & 0.867 &  0.400 &  0.000 & -0.694  \tabularnewline
 0.800 &  0.600 & 0.867 &  0.400 &  0.000 & -0.463  \tabularnewline
 0.867 &  0.723 & 0.867 &  0.400 &  0.000 & -0.232  \tabularnewline
 0.933 &  0.867 & 0.867 &  0.400 &  0.000 &  0.000  \tabularnewline
 1.000 &  1.000 & 1.000 &  0.400 &  0.000 &  0.000  \tabularnewline
\end{tabular} 
\end{ruledtabular}
\end{table}

\begin{table}[p]
  \caption{\label{tab:naphtol-ilog-schedule} Alchemical schedule of the softplus perturbation function for the hydration of 1-naphthol.}
\begin{ruledtabular}
\begin{tabular}{lccccc}
  $\lambda$  &   $\lambda_1$ & $\lambda_2$ & $\alpha$\footnote{kcal/mol$^{-1}$} & $u_0$\footnote{kcal/mol} & $w_0$\footnote{kcal/mol}  \tabularnewline
 0.000 &  0.000 & 0.000 &  0.400 &  5.000 &  0.000  \tabularnewline
 0.067 &  0.000 & 0.116 &  0.400 &  5.000 & -0.778  \tabularnewline
 0.133 &  0.000 & 0.231 &  0.400 &  5.000 & -1.557  \tabularnewline
 0.200 &  0.000 & 0.347 &  0.400 &  5.000 & -2.335  \tabularnewline
 0.267 &  0.000 & 0.462 &  0.400 &  5.000 & -3.113  \tabularnewline
 0.333 &  0.000 & 0.578 &  0.400 &  5.000 & -3.892  \tabularnewline
 0.400 &  0.000 & 0.867 &  0.400 &  5.000 & -5.837  \tabularnewline
 0.467 &  0.000 & 0.867 &  0.400 &  1.667 & -2.947  \tabularnewline
 0.533 &  0.067 & 0.867 &  0.400 &  0.000 & -1.387  \tabularnewline
 0.600 &  0.200 & 0.867 &  0.400 &  0.000 & -1.156  \tabularnewline
 0.667 &  0.333 & 0.867 &  0.400 &  0.000 & -0.925  \tabularnewline
 0.733 &  0.467 & 0.867 &  0.400 &  0.000 & -0.694  \tabularnewline
 0.800 &  0.600 & 0.867 &  0.400 &  0.000 & -0.463  \tabularnewline
 0.867 &  0.733 & 0.867 &  0.400 &  0.000 & -0.232  \tabularnewline
 0.933 &  0.867 & 0.867 &  0.400 &  0.000 &  0.000  \tabularnewline
 1.000 &  1.000 & 1.000 &  0.400 &  0.000 &  0.000  \tabularnewline
\end{tabular} 
\end{ruledtabular}
\end{table}

\begin{table}[p]
  \caption{\label{tab:diphenyltoluene-ilog-schedule} Alchemical schedule of the softplus perturbation function for the hydration of 3,4-diphenyltoluene.}
\begin{ruledtabular}
\begin{tabular}{lccccc}
  $\lambda$  &   $\lambda_1$ & $\lambda_2$ & $\alpha$\footnote{kcal/mol$^{-1}$} & $u_0$\footnote{kcal/mol} & $w_0$\footnote{kcal/mol}  \tabularnewline
 0.000 &  0.000 & 0.000 &  0.400 &  4.800 &  0.000  \tabularnewline
 0.067 &  0.000 & 0.131 &  0.400 &  4.800 & -0.212  \tabularnewline
 0.133 &  0.000 & 0.262 &  0.400 &  4.800 & -1.272  \tabularnewline
 0.200 &  0.000 & 0.392 &  0.400 &  4.800 & -1.908  \tabularnewline
 0.267 &  0.000 & 0.524 &  0.400 &  4.800 & -2.120  \tabularnewline
 0.333 &  0.000 & 0.655 &  0.400 &  4.800 & -2.544  \tabularnewline
 0.400 &  0.000 & 0.786 &  0.400 &  4.400 & -3.392  \tabularnewline
 0.467 &  0.000 & 0.917 &  0.400 &  4.400 & -3.180  \tabularnewline
 0.533 &  0.050 & 0.917 &  0.400 &  4.400 & -2.968  \tabularnewline
 0.600 &  0.156 & 0.917 &  0.400 &  4.400 & -2.756  \tabularnewline
 0.667 &  0.308 & 0.917 &  0.400 &  4.000 & -2.332  \tabularnewline
 0.733 &  0.460 & 0.917 &  0.400 &  4.000 & -1.908  \tabularnewline
 0.800 &  0.612 & 0.917 &  0.400 &  0.000 & -1.484  \tabularnewline
 0.867 &  0.764 & 0.917 &  0.400 &  0.000 & -1.060  \tabularnewline
 0.933 &  0.917 & 0.917 &  0.400 &  0.000 &  0.000  \tabularnewline
 1.000 &  1.000 & 1.000 &  0.400 &  0.000 &  0.000  \tabularnewline
\end{tabular} 
\end{ruledtabular}
\end{table}


\end{document}